\documentclass{scrartcl}
\usepackage[all]{nowidow}
\usepackage{hyperref}
\usepackage{url}
\usepackage{times}
\usepackage{xcolor}
\usepackage{booktabs} 
\usepackage{graphicx} 
\usepackage{subcaption} 


\usepackage{verbatim}
\usepackage[utf8]{inputenc}
\usepackage[T1]{fontenc}
\usepackage{subcaption}
\usepackage{amsmath}

\title{Health Estimate Differences Between Six Independent Web Surveys: Different Web Surveys, Different Results?}

\author
{Rainer Schnell$^{1\ast}$ and Jonas Klingwort$^{2}$\\
\\
\normalsize{$^{1}$Research Methodology Group, University of
    Duisburg-Essen, 47057 Duisburg, Germany} \\
\normalsize{$^{2}$Department of Research \& Development, Statistics Netherlands (CBS), CBS-weg 11,}\\ \normalsize{PO Box 4481, 6401 CZ Heerlen, the Netherlands}
~ \\[2mm]
\normalsize{$^\ast$rainer.schnell@uni-due.de}}
    \date{}

\begin{document} 
    
    \maketitle 
    \begin{abstract}
        \textbf{Abstract:}
Most general population web surveys are based on online panels maintained by commercial survey agencies. However, survey agencies differ in their  panel selection and management strategies. Little is known if these different strategies cause differences in survey estimates. This paper presents the results of a systematic study designed to analyze the differences in web survey results between agencies. Six different survey agencies were commissioned with the same web survey using an identical standardized questionnaire covering factual health items. Five surveys were fielded at the same time. A calibration approach was used to control the effect of demographics on the outcome. Overall, the results show differences between probability and non-probability surveys in health estimates, which were reduced but not eliminated by weighting. Furthermore, the differences between non-probability surveys before and after weighting are larger than expected between random samples from the same population.
    \end{abstract}
    
    
    \section{Introduction}

Web surveys are increasingly used for public health research \cite{Liu2010, Russell2010, Haddad2022Dec}, official statistics \cite{Klingwort2018}, social and marketing research \cite{ADM2019}, and election and opinion polls \cite{Sohlberg2017Oct, Sturgis2018Jun}. Due to the low costs, short fieldwork periods, the ease of elaborate filtering, and the option to use multimedia elements in questionnaires \cite{BlairSzajaBlair2014}, the popularity of web surveys is plausible. Furthermore, the absence of interviewers often seems to be an argument for less socially desirable response behaviour \cite{KreuterPresserTourangeau2008}. Finally, during the Covid-19 pandemic, web surveys had the additional advantage of preventing any physical contacts.

However, the main methodological problem of web surveys is sampling. For general population surveys, with very few exceptions, no sampling frame for direct contacts to respondents (e.g., email addresses) is available. Population covering sampling frames containing email addresses are commonly restricted to special populations, such as specific professions. Therefore, most commercial web surveys are based on non-probability samples, usually recruited online. In general, online recruitment of new panel members uses, for example, social networks, online communities/social media, affiliate networks, or website banners \cite{AAPOR2023}.

Since most web surveys are based on non-probability samples, differences between differently recruited surveys can be expected \cite{Cornesse2020Feb}. Because probability samples are based on easily reproducible procedures with error bounds given by sampling error, differences between probability samples should be small. In contrast, the procedures for non-probability samples are hardly described in detail, making reproduction difficult. Therefore, we expect larger differences between non-probability samples and probability samples, as could be expected between different random samples from the same population.

There is a lack of empirical studies evaluating whether different agencies will produce similar results for the same questionnaire on factual items using web surveys. A recent study by \cite{Pekari2022Dec}, for example, compares demographics and voting outcomes, but no other factual items, between two random samples samples and three non-probability samples.

This article presents the results of the first systematic empirical study of potential differences between health-related web surveys. Therefore, six independent web surveys were commissioned. Four agencies used non-probability samples, two agencies probability-based samples.

\section{Research background}\label{sec:background}

\subsection{Undercoverage and nonresponse in web surveys}\label{subsec:coverage}

Currently, due to the absence of a sampling frame of the general population, random sampling for single-mode web surveys is impossible \cite{Couper2000} under almost all jurisdictions. Although other sampling procedures are possible, in practice, web surveys are often based on online-recruited access panels, or participants are recruited on websites visited. Therefore, elements are missing in the sample due to coverage errors, and inclusion probabilities are unknown for those responding. Accordingly, design-based estimates of web surveys are not legitimated by mathematical statistics \cite{Bethlehem2015a}. Some survey agencies recruit offline by drawing random samples, for example from a population register and inviting sampled elements to participate in a web survey to remedy this problem. Furthermore, if persons without internet access are provided with online access, recruiting offline may reduce undercoverage bias \cite{Leenheer2013, Blom2016Jun, Cornesse2021}.  

Although the level of internet access at the European Union's household level has increased steadily, differences still exist:  from 85\% in Greece to 99\% in Norway (in 2022) \cite{Eurostat2023a}. For Germany in 2022, official statistics reported internet access at the household level at 91\% and individual internet usage at about 93\% \cite{Eurostat2023a,Eurostat2023b}. Based on the American Community Survey 2021, the internet penetration rate in the USA is estimated at 90\% \cite{uscensus2023}. Depending on the excluded proportion of people without internet access and the difference in the target variable between people with internet and without internet \cite{BethlehemBiffignandi2012}, excluding subpopulations might bias survey estimates.

While undercoverage addresses the internet access requirement, nonresponse refers to the ability and motivation to participate in the survey. Both selection mechanisms, undercoverage and nonresponse, may cause bias in the survey estimates. Accordingly, a  survey data set is the result of both selection mechanisms. Disentangling between coverage and nonresponse errors is impossible due to the absence of a sampling frame \cite{AAPOR2010}. The larger the proportion of non-respondents and the stronger the correlation between the target variable and the missing data mechanism, the larger the nonresponse bias. 

An equation by \cite{BethlehemBiffignandi2012} allows the estimation of the difference between a population mean $\bar{Y}$ and a mean of a sample with nonreponse $\bar{Y}_{nr}$:

\begin{equation}
\bar{Y} - \bar{Y}_{nr} \approx \frac{R_{\rho Y} S_{\!\rho} S_{\!Y}}{\bar{\rho}}.
\label{eq:bias}
\end{equation}

The equation (\ref{eq:bias}) assumes that every person has a response propensity $\rho$, with an overall mean $\bar{\rho}$, and standard deviation $S_{\rho}$ in the population. $R_{\rho Y}$ is the correlation between $Y$ and $\rho$, and $S_{Y}$ is the standard deviation of $Y$. The bias ($\bar{Y} - \bar{Y}_{nr}$) depends on three quantities: the correlation between the response propensity and the variable to be estimated, the variance of the response propensity, and the variance of the variable of interest. Accordingly, the bias will be small if the participation rate in the non-probability sample is high or $R_{\rho Y}$ is small or $S_{Y}$ is small.

Regardless of response mode or mandatory participation, surveys show a downward trend in response rates \cite{Meyer2015, Czajka2016, Williams2017, Leeuw2018}. Web surveys yield even lower response rates than other response modes \cite{Daikeler2020}. In general, with increasing proportions of nonresponse and increasing correlations between the response variable and the mechanisms causing nonresponse, the risk for biased estimates increases \cite{Groves2008Jan}. However, although probability-based surveys suffer from decreasing response rates, empirically, their estimates seem to be still more accurate than estimates obtained by non-probability samples \cite{Cornesse2020Feb}. Given equation (\ref{eq:bias}), this empirical result is mathematically plausible only if the correlation between response propensity and variables of interest is low in probability-based surveys.

\subsection{Internet use and health}\label{subsec:internethealth}

The mechanisms causing differences in internet use depending on health conditions are rarely discussed. However, the selection process from the target population to the sampled population of web surveys can be summarized in six steps. First, the technical requirements of a working internet connection by line, WiFi, digital cellular networks or satellites must be fulfilled. Second, sufficient financial resources by the respondents are necessary if internet access is not provided for free. Third, using a smartphone or a computer requires the physical ability to see (or hear) and the ability to type or speak. Fourth, answering survey questions requires cognitive abilities such as understanding abstract concepts, word finding and judgment. Fifth, recruitment for a survey needs a mode to contact the respondent, which usually requires a sampling frame. Such a frame for web-based population surveys is rare. Therefore, offline sampling requires other contact modes, such as phone numbers or address lists. For online sampling, river sampling or similar non-probability sampling techniques are necessary. Sixth, the designated respondent needs sufficient motivation to answer a survey request.

Physical or cognitive capabilities might be impaired depending on the symptoms of a medical condition. Due to hospital stays or caregivers, the probability of contact with the designated respondent may vary between contact modes and medical conditions. Finally, motivation for a response might decrease (or increase) depending on the type and severity of the medical condition.

The effects might not necessarily be linear or additive. For example, physical disabilities might impact survey participation only for severely disabled persons. Increasing physical or cognitive problems might reduce motivation as well. Therefore,  neither a diagnosis (a reported ICD number) nor a specific symptom alone will be a sufficient or necessary condition for survey response. Hence, no simple pattern of health conditions and survey participation can be expected.

However, some studies are available if the potential bias could be reduced by weighting. \cite{AdamsWhite2007} reported that weighting by age and gender did not eliminate differences between a web and a CAPI survey in BMI, eating habits, physical activity, alcohol consumption, and smoking. \cite{Russell2010} reported that web responders currently smoked less, had fewer children, and less often had a chronic disease. Observed health differences between internet users and non-users (based on the Michigan Behavioral Risk Factor Surveillance System (BRFSS) survey 2003 and the Health and Retirement Study 2002) are described by \cite{TourangeauFrederickCouper2013}. Based on reported internet usage in European (European Social Survey) and US data (BRFSS) obtained by conventional surveys (F2F and CATI), \cite{Schnell2017} note that `(...) calibration on age, gender, ethnic background, urban residence, education and household income does not eliminate the observed health differences'. In a probability-based survey with the web as a response mode, \cite{Braekman2020Jan} showed significant differences between web and face-to-face respondents after controlling for gender, age, region, marital status, household size, educational attainment, and country of birth. Recently, \cite{Dutwin2022May} showed in a comprehensive analysis of American data persisting differences between internet users and non-users including age, employment, cultural activities, and education. Using British and Swedish data, \cite{HelsperReisdorf2017} provided evidence for similar differences regarding, for example, age, low-level of education, and living alone. Both publications noticed health issues (such as disability) as predictors for internet use. \cite{Dutwin2022May} summarized their findings: `Without some reasonable adjustment, a variable like health status has a high risk of being significantly biased in studies that do not cover the non-internet population'. Hence, there is growing evidence of correlations between health indicators and internet use, which cannot be corrected by weighting procedures. 

\subsection{Selection mechanisms and weighting}

The data missing due to coverage and nonresponse errors can be described by three different data generating mechanisms: missing completely at random (MCAR), missing at random (MAR), and missing not at random (MNAR) \cite{ZhouZhouLiuDing2014, LittleRubin2020}. 

If the data generating mechanism is MAR, the generalized regression estimator (GREG) can be used to correct the effect of such missing data by calibration \cite{SaerndalSwenssonWretman1992}. The calibration estimator for respondents $(r)$ of a target variable $Y$ is defined as

\begin{equation}
    \hat{Y}_w = \sum_r w_i y_i,
\end{equation}

with $w_i$ the calibrated correction weights and $y_i$ the response. The $w_i$ are a product of the initial weights $wi_i$ and the correction factor $v_i$, where $w_i = wi_i*v_i$ (for details on the calibration estimator, see \cite{SaerndalLundstroem2005}). When the MAR assumption does not hold, GREG estimates are still biased. In such cases, the probability of being included in the survey and response depends on $y_i$, and $x_i$ cannot fully explain the selection effect. Therefore, the missing data generating mechanism is considered as MNAR. 

\section{Survey agencies and recruitment strategies}\label{subsec:surveyagencies}

The research design of the study presented here intended to commission five different survey agencies to collect health data using their online panels. The agencies included are the largest commercial market research companies offering their panels for academic research in Germany. However, since none of these panels were based on probability samples, the only commercial probability-based web panel was also included. Due to an additional grant, a sixth web survey could be conducted five months after the last interview of the other five surveys. This panel is the only general purpose academic probability-based web survey in the country under study. The target population was defined as the residential population aged 18 years and older. Details of the fieldwork are shown in Table \ref{tab:samples}.

\begin{table}[htbp]
\centering
\caption{Agency, sampling type, recruitment, fieldwork duration, and sample size of the six surveys.}
\label{tab:samples}
\begin{tabular}{llccc}
\toprule
Agency& \multicolumn{1}{c}{Sampling} & \multicolumn{1}{c}{Recruitment} & Fieldwork period & Sample size \\
\midrule
NPS-$1$ & Quota sample  & online  & 13.09. -- 30.09.2016 & 5.002 \\ 
NPS-$2$ & Quota sample  & online  & 13.09. -- 28.09.2016 & 5.000 \\ 
NPS-$3$ & Quota sample  & online  & 13.09. -- 16.11.2016 & 5.501 \\ 
NPS-$4$ & Quota sample  & online  & 13.09. -- 19.09.2016 & 2.877 \\ 
PS-$1$  & Random sample & offline & 13.09. -- 07.10.2016 & 5.001 \\ 
PS-$2$  & Random sample & offline & 19.04. -- 13.06.2017 & 3.065 \\ 
\bottomrule
\end{tabular}
\end{table}

NPS-$1$, NPS-$2$, and NPS-$3$ are globally operating providers of online panels.\footnote{NPS-1 is operated by GMI (now part of Kantar), NPS-2 is operated by SSI (now named Dynata), NPS-3 is operated by Ipsos, NPS-4 is the WiSo-panel \cite{Goeritz2014}, PS-1 is operated by Forsa, and PS-2 is operated by GESIS.}
These agencies use opt-in access panels where participants have deliberately registered. The commissioned agencies also used quota sampling to approximate the population demographics with their samples. NPS-$4$ is an ongoing online opt-in panel managed by a university. This panel invited all members of the panel to obtain a sample for this study. 

PS-$1$ and PS-$2$ used probability samples. PS-$1$ is a commercial research company, PS-$2$ a publicly financed research agency. While survey agency PS-$1$ used a random-digit dialing probability sample \cite{Guellner2004}, PS-$2$ used F2F interviews of a register-based sample to recruit members for panel-participation \cite{Cornesse2021}. Both PS-surveys provided internet access to previously offline respondents. 

All agencies were contractually obliged to implement the same survey, with a questionnaire developed and tested in advance. However, all details of the fieldwork were left to the agencies to reflect their actual practice. \footnote{Therefore, many different panel management strategies could impact differences between agencies, for example, recruitment, payment, control and web interface. Furthermore, providers may have different panel attrition problems or suffer from different panel conditioning effects. Separating these effects could form a research program on its own.}

To control for time-dependent influences, the initially planned five surveys started fieldwork at the same time. The sample size to be delivered was contractually set at $n=5.000$. Agencies with smaller panels (NPS-$4$ and PS-$2$) delivered smaller sample sizes. No further data pre-processing was applied. Hence the data delivered by the agencies were analyzed. To sum up: NPS-$1$, NPS-$2$, NPS-$3$, NPS-$4$ used undefined mixtures of non probability samples. PS-$1$ and PS-$2$ used probability samples.

\subsection{Questionnaire and pretest}\label{subsec:questionnaire}

Health was chosen as the survey topic because undercoverage and nonreponse due to health issues seem likely given previous research (see Section \ref{subsec:internethealth}). The questionnaire contained non-attitudinal questions on general health status, health-care utilization, injuries and accidents, disabilities and chronic diseases, and health-related behaviour. In total, 36 health items were used. The majority of items were taken from three general population health monitoring studies (DEGS-1 \cite{GoesswaldLangeDoelleHoelling2013, Kamtsiuris2013Die}, GEDA-12 \cite{KochInstitut2014}, and GEDA-14/15 \cite{Sass2017}).

Six items on demographics were asked to be used in the weighting model. The questions on age and municipality size were taken from a reference survey \cite{ALLBUS2014}. The questions on education were adopted from official statistics \cite{StatistischesBundesamt2010}. The questions on gender and federal state were developed in collaboration with the survey agency that conducted the pretest.

The questionnaire was pretested using an online recruited access-panel of a market research agency. In total, 550 respondents of a quota sample tested the questionnaire. Only seven aborted interviews, not clustered at specific questions, were identified. No question produced implausible responses or large proportions of item-nonresponse. The questionnaire required, on average, about 5 minutes to complete. 

\subsection{Dependent variables}\label{sec:vars}

In total, 36 health items will be used as dependent variables in the analysis. A dichotomized five-point ordered subjective health indicator (good/very good: 1; else: 0), and the Body-Mass-index (calculated from questions on weight and height) were used for general health status. Six questions asked for the use of health services:

\begin{enumerate}
    \item the number of general practitioner visits in the last four weeks, 
    \item the number of general practitioner visits in the last 12 months, 
    \item the number of calendar days being ill and unable to perform usual duties in the last 12 months, 
    \item the number of working days being diagnosed by a physician as unable to work within the last 12 months,
    \item the number of overnight hospital stays for inpatient treatment within the last 12 months and 
    \item  if an artificial hip joint replacement was implemented in the last 12 months.
\end{enumerate}

For the latter question as for all dichotomous questions, the code 0 indicates 'no', code 1 indicates 'yes'. 

Four dichotomous indicators were asked on traffic accidents, accidents at home, accidents in leisure time, and injuries at work in the last 12 months. 15 items related to ever diagnosed diseases (high blood pressure, allergy, chronic back pain, sleep disorder, joint diseases (arthrosis/rheumatism), depressive disorder, migraine, heart diseases (heart failure/cardiac insufficiency), chronic bronchitis, diabetes, osteoporosis, liver diseases (fatty liver/liver infection/hepatitis/liver shrinkage/cirrhosis), asthma, stroke, and cancer. 
 
Further, the respondents were asked to indicate their use of glasses/contact lenses, their ability to read a newspaper without difficulty and their use of hearings aids. Finally, they should report if a disability condition was officially certified, and if so, which degree of disability (20-100) was certified.

Finally, four items covered health-related behaviour. Smoking was assessed by asking for the number of smoked cigarettes per day and week. Sports activity was dichotomized (less than one hour/at least one hour per week). Drinking alcoholic beverages was also dichotomized (less than four days or at least four days per week). However, due to a restrictive interpretation of German data protection law by PS-2, only 14 out of 36 variables are provided for the PS-2 data set.\footnote{The data would have been available in a  closely supervised research data center, but initially, PS-$2$ was not able to grant access within six months to the research data center. Later, Covid-19 restrictions delayed access to the research data center.}

\section{Methods}\label{sec:methods}

\subsection{Methods for comparisons}\label{sec:methodscompare}

Data was analyzed in four steps: Beginning with unweighted survey estimates, ANOVA was used for testing mean differences in unweighted estimates between surveys. In the second step, Generalized Regression Estimation (GREG) was used for weighting the surveys. In the third step, multiple pairwise comparisons between the weighted means of surveys using Tukey’s Honest Significant Difference Test (Tukey’s HSD) were computed. Finally, we pooled the non-probability samples into a single NPS group to reduce the number of comparisons and use t-tests for comparison.\footnote{Since two heterogeneous kinds of samples have to be compared, we have no meta-analysis problem, which excludes standard measures of heterogeneity. Therefore, we use multiple pairwise comparisons (Tukey's HSD) between the weighted means of surveys.}

Since the comparisons for 14 variables are based on the comparisons between all six web surveys, we have $(6 \times 5) / 2 = 15$ pairwise comparisons, in total $15 \times 14 = 210$ comparisons for six surveys. In addition, we have 22 variables for five surveys, giving $(5 \times 4) / 2 = 10$ pairwise comparisons of 22 variables giving $22 \times 10 = 220$. Overall, we have $210 + 220 = 430$ group comparisons.

To simplify the results by reducing the number of comparisons  we pooled the non-probability samples into a single NPS group in a separate analysis. For these comparisons between PS and NPS samples, t-tests were used. For PS-1 versus NPS, 36 items are available for comparison and for PS-2 versus NPS, 14 items. We first compare unweighted estimates and then weighted estimates. Therefore, in total, there are $k = 2 \times 36 + 2 \times 14 = 100$ comparisons.\footnote{Comparing p-values with a fixed threshold is rarely advised \cite{Wasserstein2019}. We use t-tests here as rough indicators for differences larger than expected, not to make decisions about a hypothesis. However, the effect measure Cohen $d$ is related to $t$:  $(|t|=\sqrt{\left(n_1 n_2\right) /\left(n_1+n_2\right)} d)$ \cite{Flury1986Aug}. The factor for multiplying $d$ to yield $t$ is about 38.7 and 50 for all comparisons. Due to this linear transformation, an analysis based on $d$ would, therefore, yield comparable results. To help interpreting the results, we additionally report effect sizes using Cohen's $d$.} To account for potential problems due to multiple testing, we apply a Bonferroni correction ($\alpha_{adj} = \frac{\alpha}{k}$), which is widely regarded as conservative \cite{Bickel2020}.

\subsection{Weighting}

Weighting adjustments for web surveys are commonly based on demographics, such as age and gender \cite{Callegaro2015, Toepoel2016}. The weighting model in the study reported here is based on demographics as well. The surveys are weighted by region, size of the municipality, age, gender, and education. Population totals of the Census were used for calibration.\footnote{Age was used with six categories (18--24, 25--29, 30--39, 40--49, 50--64, and 65+), gender with two categories, education with five categories, size of the municipality with three categories (10.000--20.000, 20.001--100.000, 100.000 and more inhabitants) and region with 16 categories (the German federal states). The GREG weighting model can be written as age~$*$~gender~$*$~education~$*$~size of municipality~$*$~federal state.} 
However, due to item-nonresponse, the information required for weighting was unavailable for some respondents.\footnote{Between 0.4\% (NPS-$2$) and 8.3\% (PS-$1$) respondents did not answer at least one question on demography.} These observations were removed from the analysis.\footnote{During the weight computations, empty cells in the weighting model were replaced with one pseudo-observation for each missing cell. The number of created pseudo-observations per survey were 1.566 for NPS-$1$, 1.628 for NPS-$2$, 1.505 for NPS-$3$, 1.965 for NPS-$4$, 1.714 for PS-$1$, and 1.839 for PS-$2$. After calculating the weights, the pseudo-observations were removed from the data set.}

\begin{figure}[htbp]
    \centering
    \includegraphics[width=.65\textwidth]{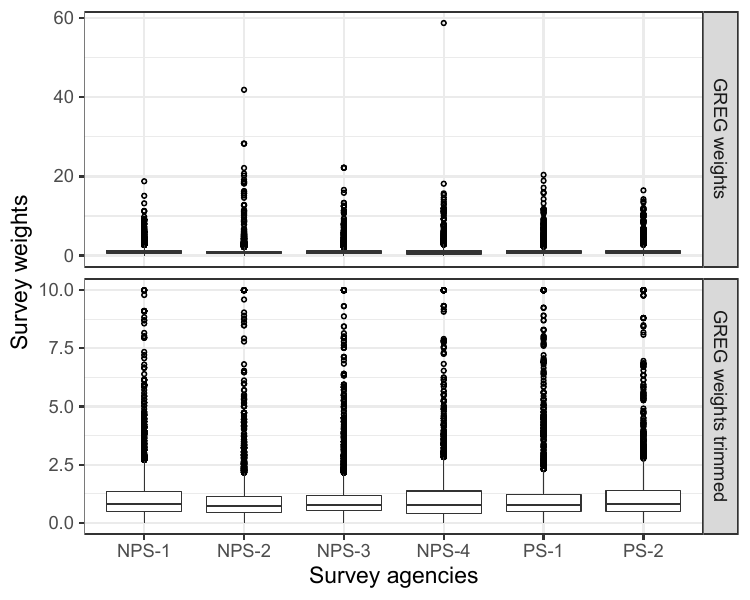}
    	\caption{Boxplots of weight distributions for each survey. The upper row shows initial survey weights, the lower row shows trimmed survey weights.}
	\label{fig:weightdistri}
\end{figure}

Figure \ref{fig:weightdistri} shows the distribution of the computed GREG weights for each survey. The upper row shows the initial weights. Especially the large weights indicate selection problems. The use of such weights would substantially increase the sampling error. In accordance to survey practice, the weights were trimmed \cite{Potter2015, ChenElliott2017}. Trimming weights will introduce a bias in the weighted estimates \cite{elliot2008}. Therefore a moderate amount of trimming was used by setting the maximum value of weights to 10. The lower row shows the distribution of the trimmed weights used for the following analysis below.

\section{Results}

\subsection{Differences between unweighted estimates}

Figure \ref{fig:naiveestimatesANOVA} shows the unweighted survey estimates and 95\%-confidence intervals for each survey. Overall, the unweighted differences between the web surveys are larger than expected by sampling alone. Regarding general health status variables, PS-$1$ has the healthiest respondents in terms of health self-assessment, and the mean BMI is also one of the lowest but still indicates a pre-obese state. The mean of respondents of NPS-$3$ and NPS-$4$ corresponds to obesity grade I.

For most variables, estimates of PS-$1$ are among the lowest for the use of health services. Respondents of NPS-$1$, NPS-$2$, and NPS-$4$ show the largest numbers of consultations.

Concerning the 15 items on ever-diagnosed diseases, the survey PS-$1$ shows the highest proportions in five variables (high blood pressure, allergy, joint diseases, heart disease, and cancer). These are mostly diseases with high incidence in the general population. Furthermore, larger proportions of PS-$1$ and PS-$2$ respondents wear glasses/contact lenses and read the print of a newspaper without problems. PS-$1$ has the largest proportion of respondents wearing hearing aids.

Furthermore, PS-$1$ and PS-$2$ show the lowest proportions of respondents with an officially certified disability. In addition, respondents of PS-$1$ and PS-$2$ are more active in sports and smoke less per day. However, on a weekly basis, respondents of PS-$2$ smoke the most. Finally, PS-$1$ and PS-$2$ show the largest proportions of respondents drinking alcoholic beverages $\geq 4$ times per week.

\begin{figure}[htbp]
\centering
	\begin{subfigure}[htbp]{1\textwidth}
	\caption*{\textit{Begin Figure 2}}
		\centering
		\includegraphics[width=0.32\linewidth]{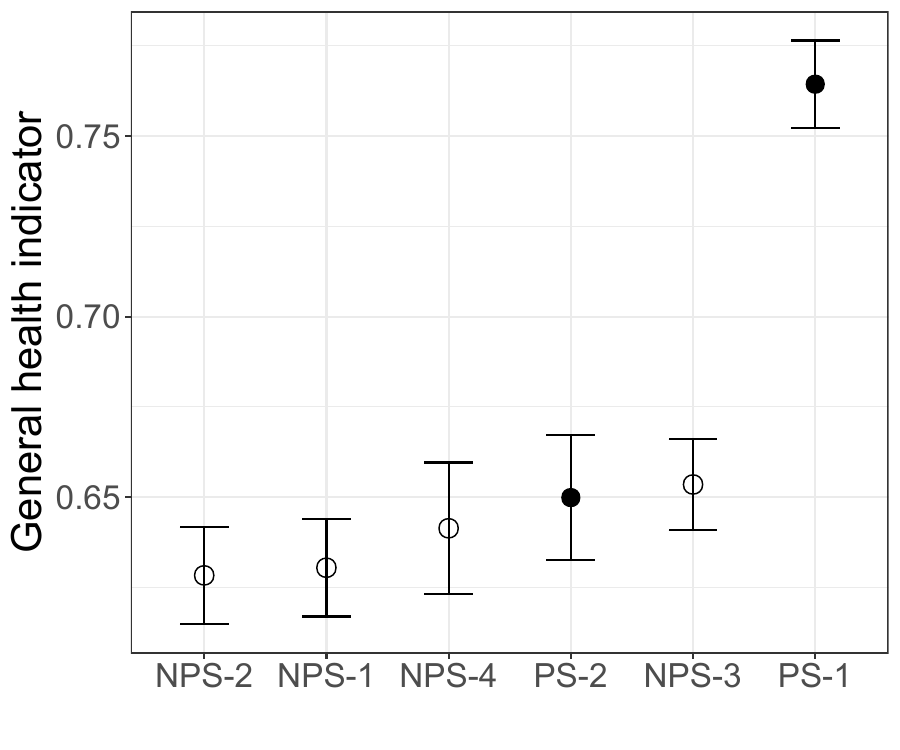}
		\includegraphics[width=0.32\linewidth]{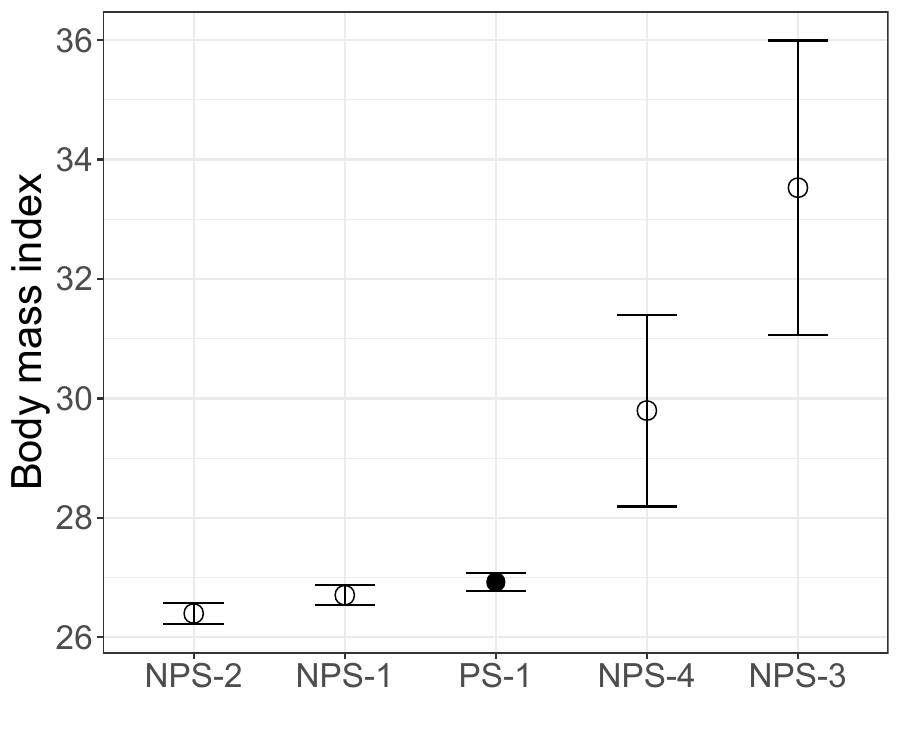}
		\includegraphics[width=0.32\linewidth]{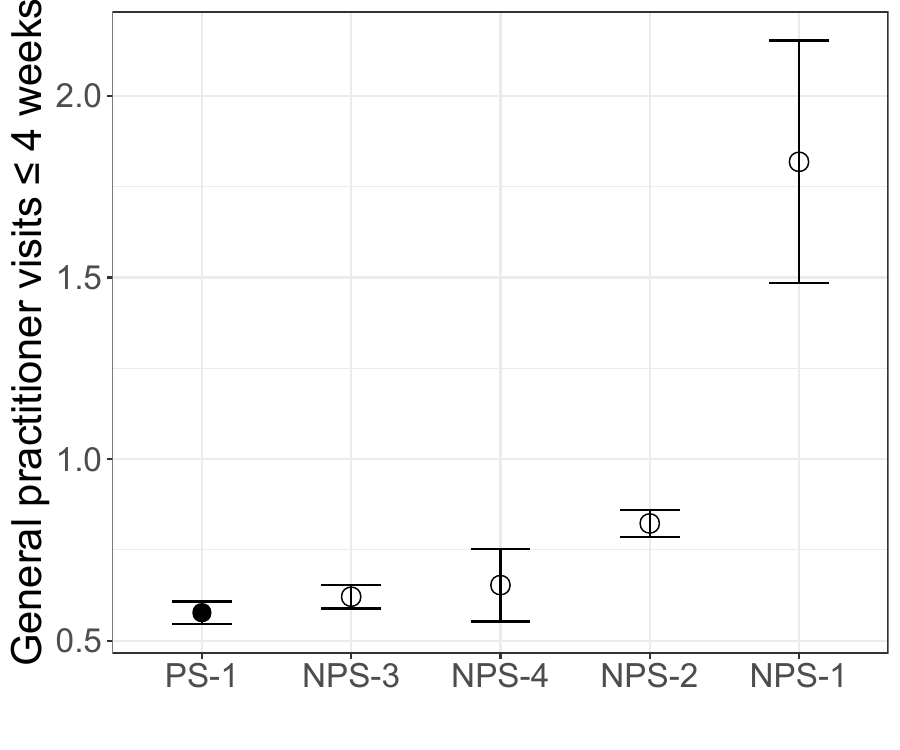}
		\includegraphics[width=0.32\linewidth]{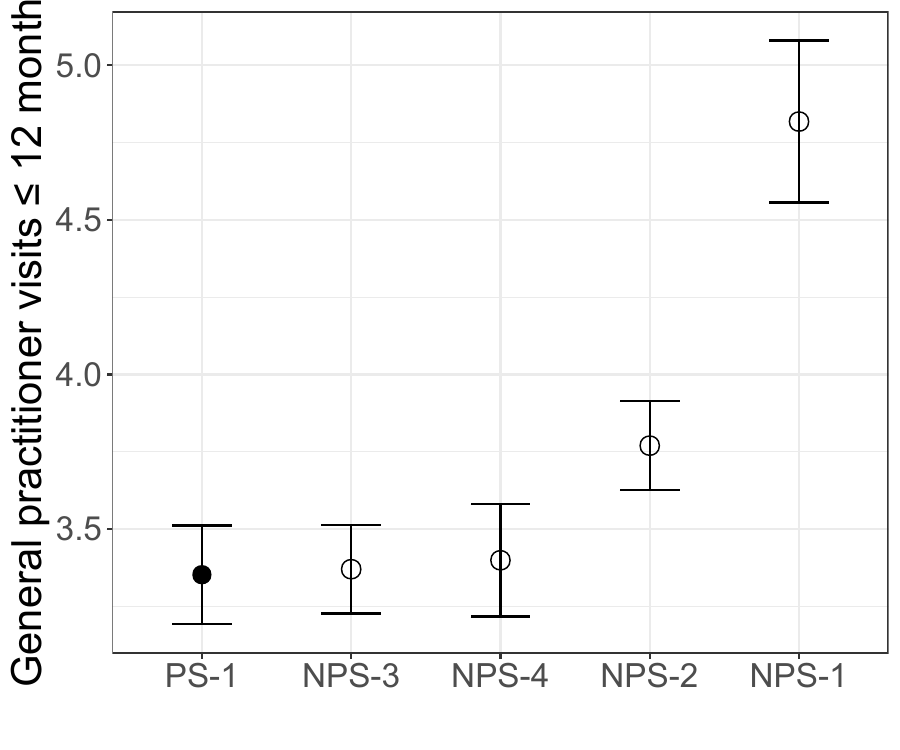}
		\includegraphics[width=0.32\linewidth]{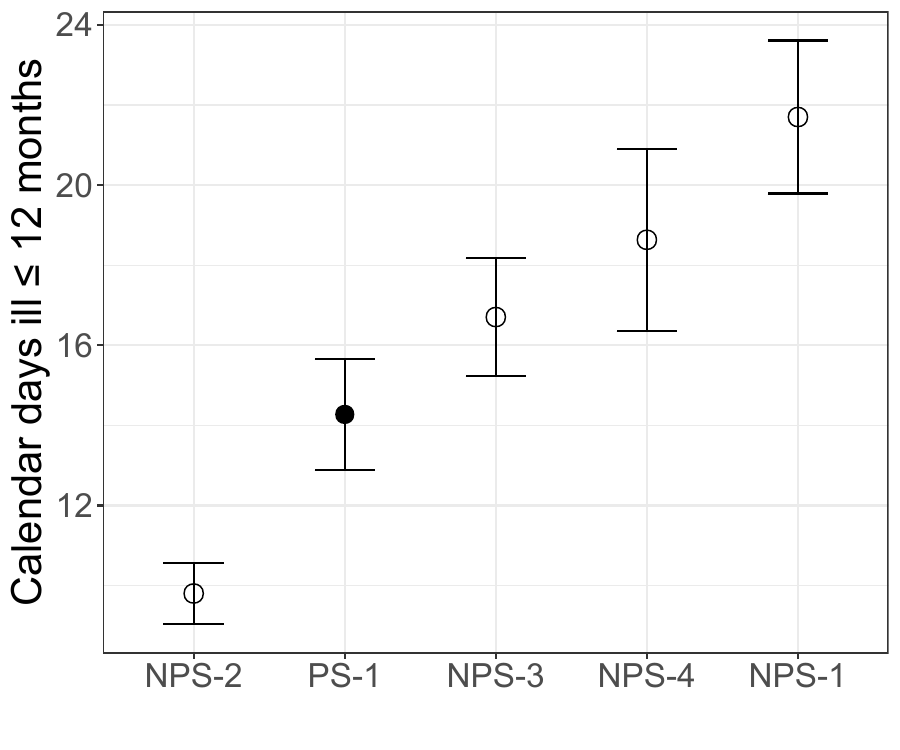}
		\includegraphics[width=0.32\linewidth]{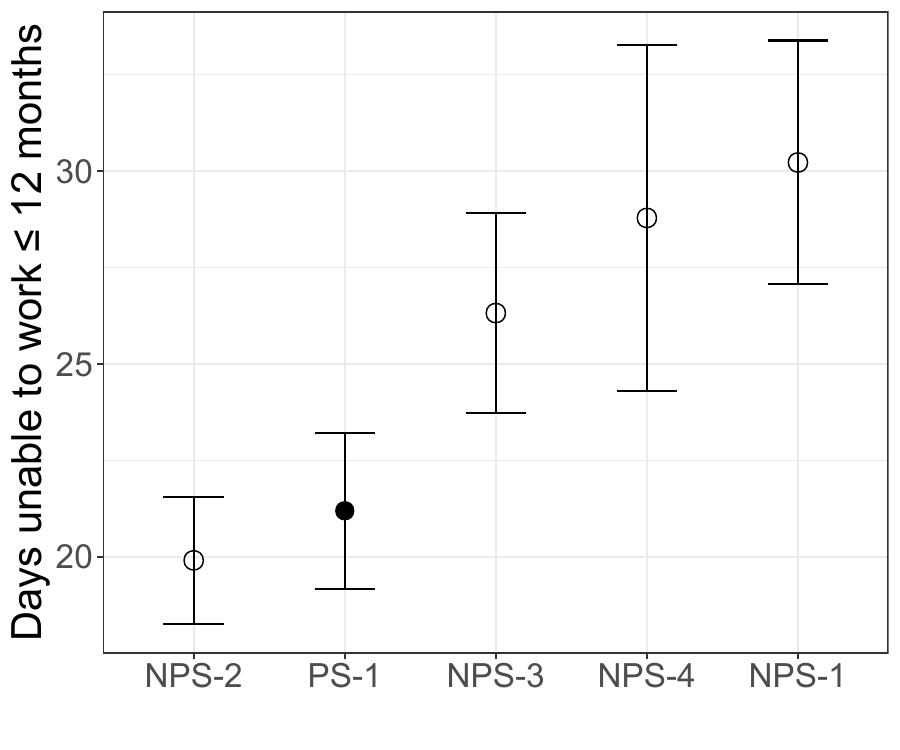}
		\includegraphics[width=0.32\linewidth]{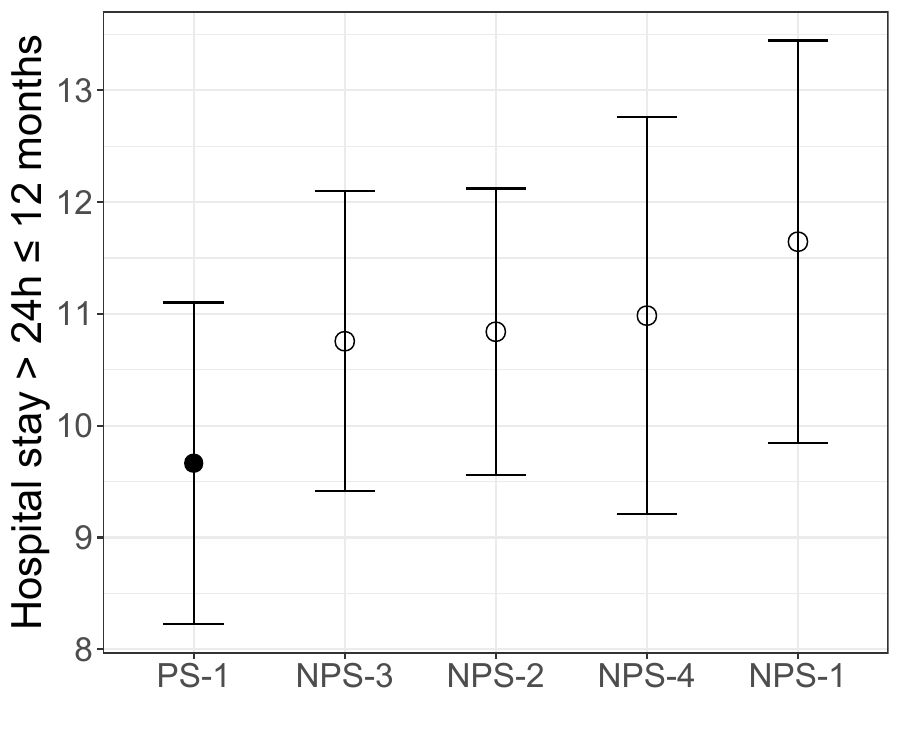}
		\includegraphics[width=0.32\linewidth]{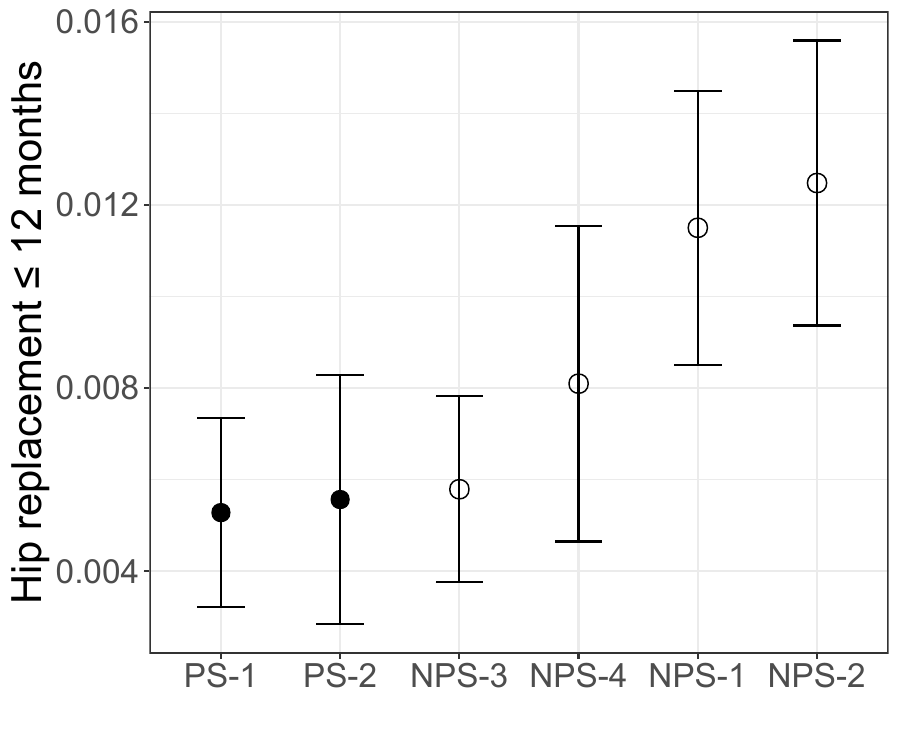}
		\includegraphics[width=0.32\linewidth]{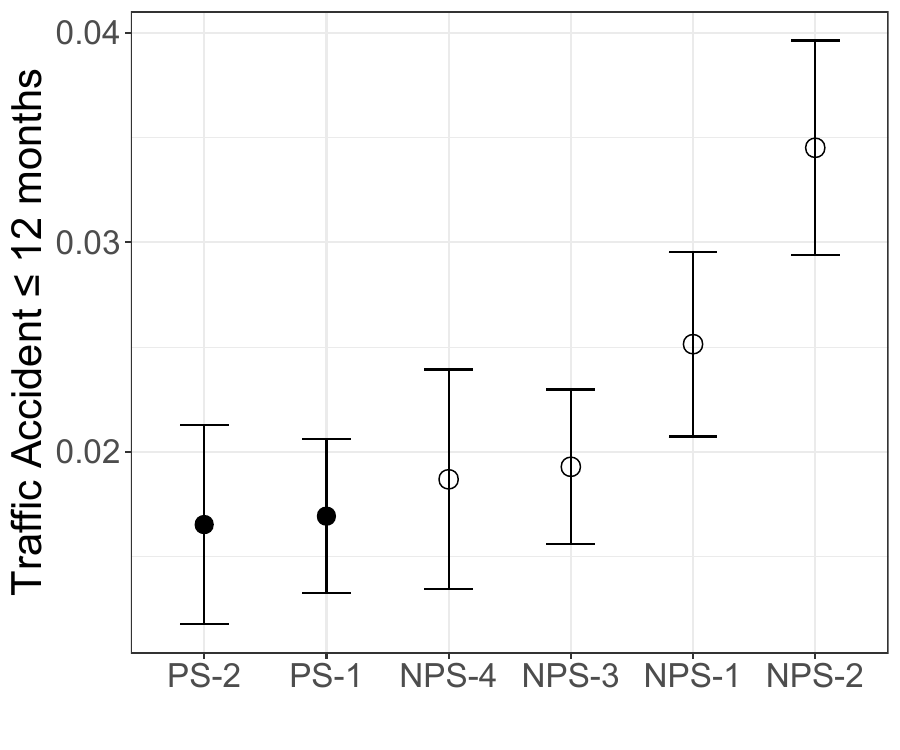}
		\includegraphics[width=0.32\linewidth]{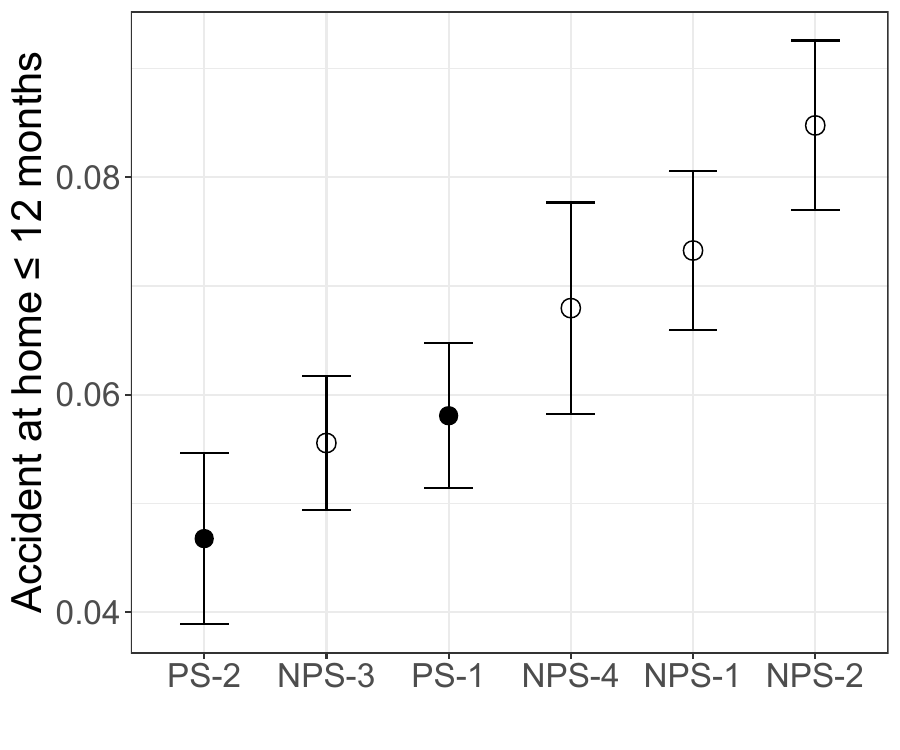}
		\includegraphics[width=0.32\linewidth]{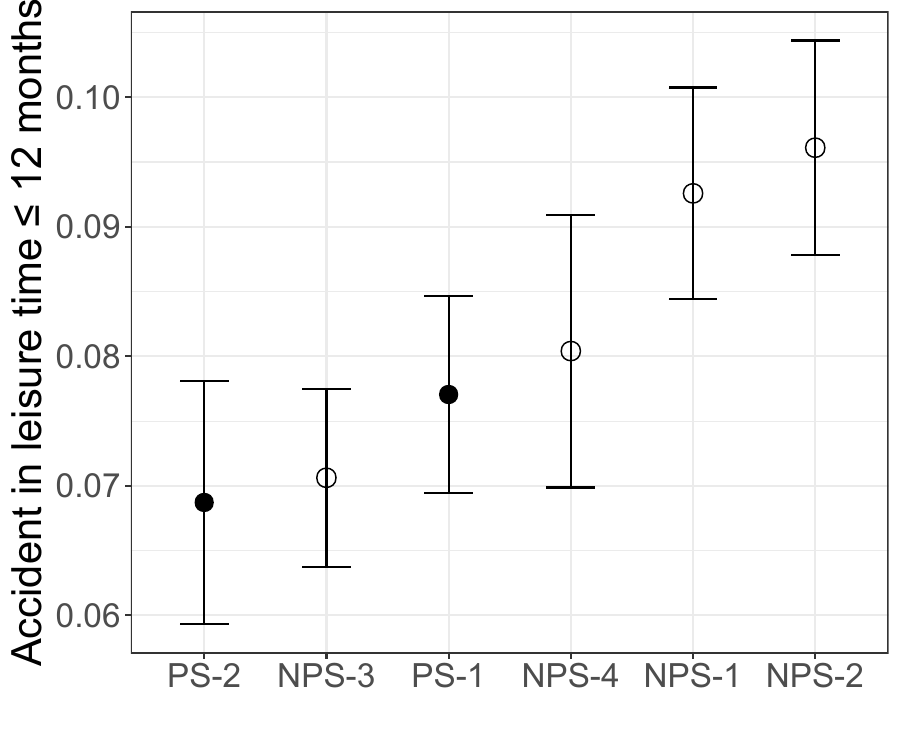}
		\includegraphics[width=0.32\linewidth]{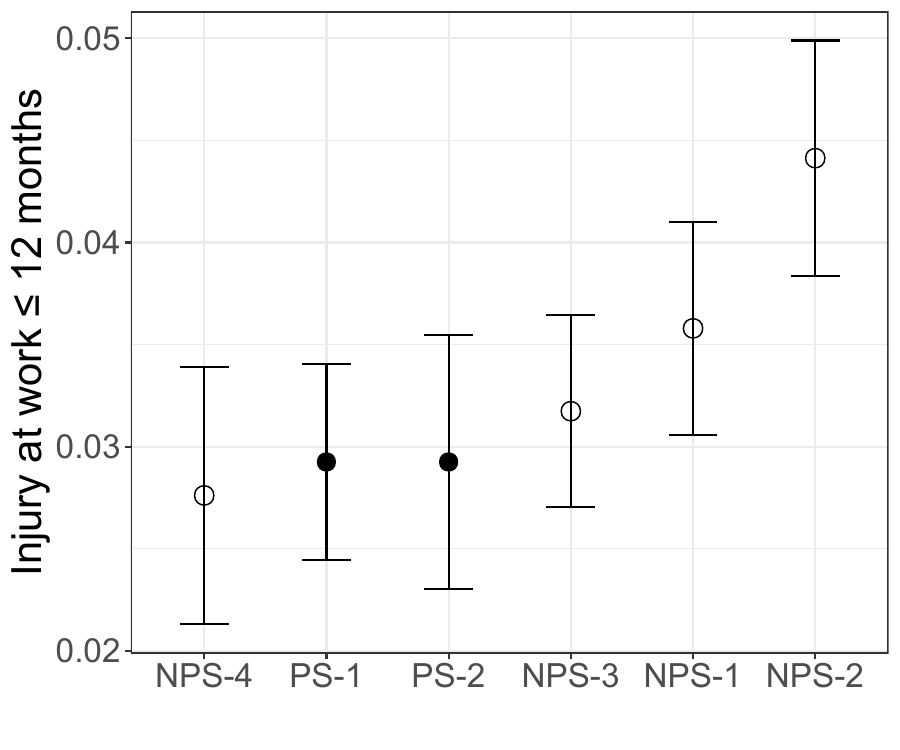}
		\includegraphics[width=0.32\linewidth]{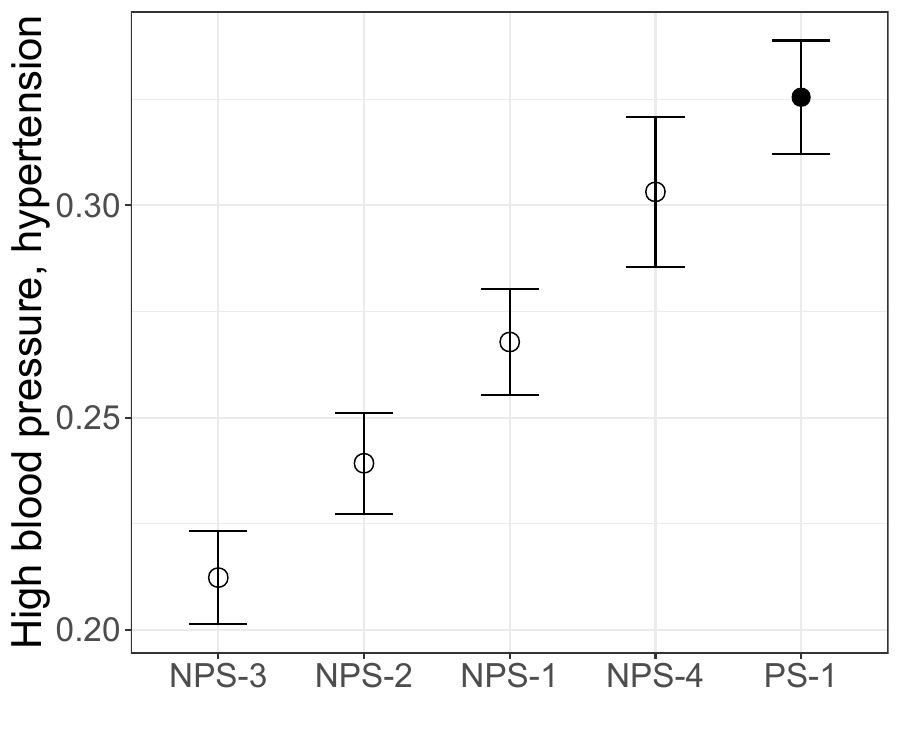}
		\includegraphics[width=0.32\linewidth]{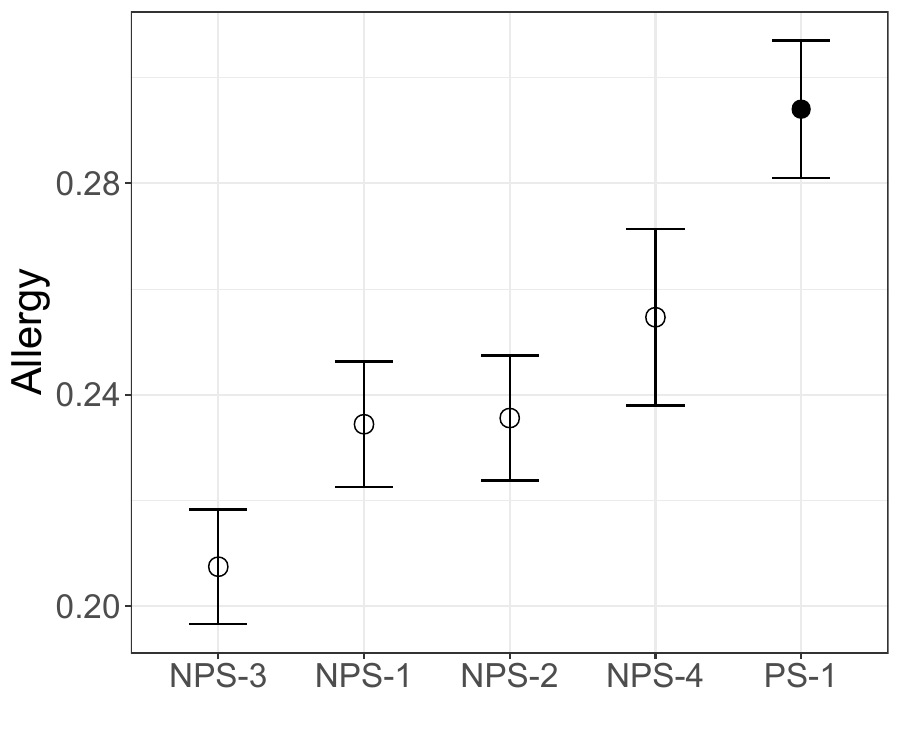}
		\includegraphics[width=0.32\linewidth]{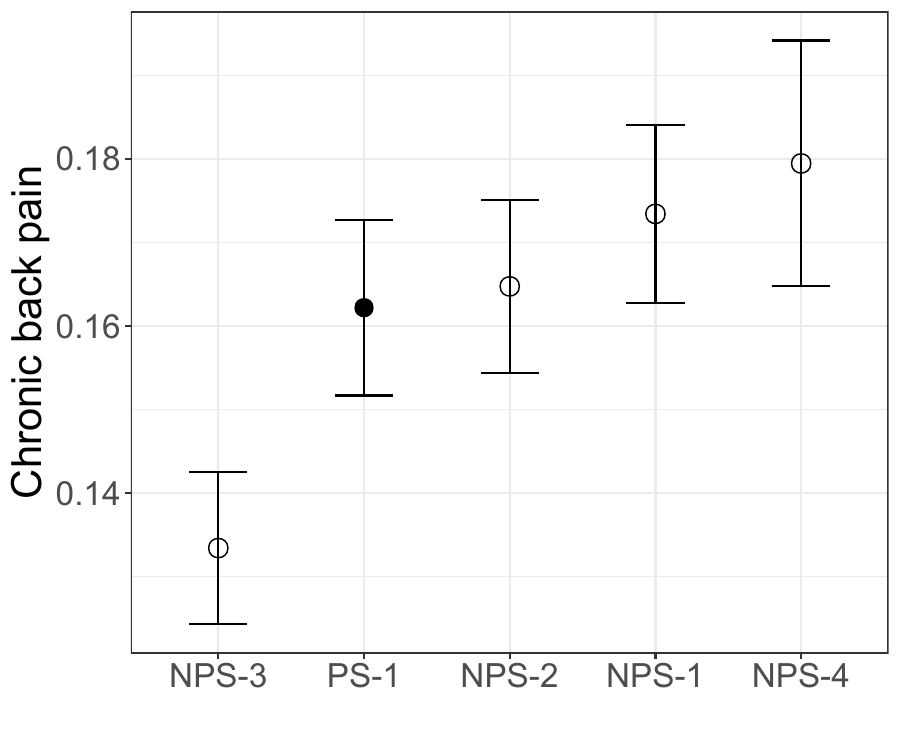}
	\end{subfigure}
\end{figure}
\begin{figure}
	\ContinuedFloat
	\begin{subfigure}[htbp]{1\textwidth}
		\caption*{\textit{Figure 2 continued}}
		\centering
		\includegraphics[width=0.32\linewidth]{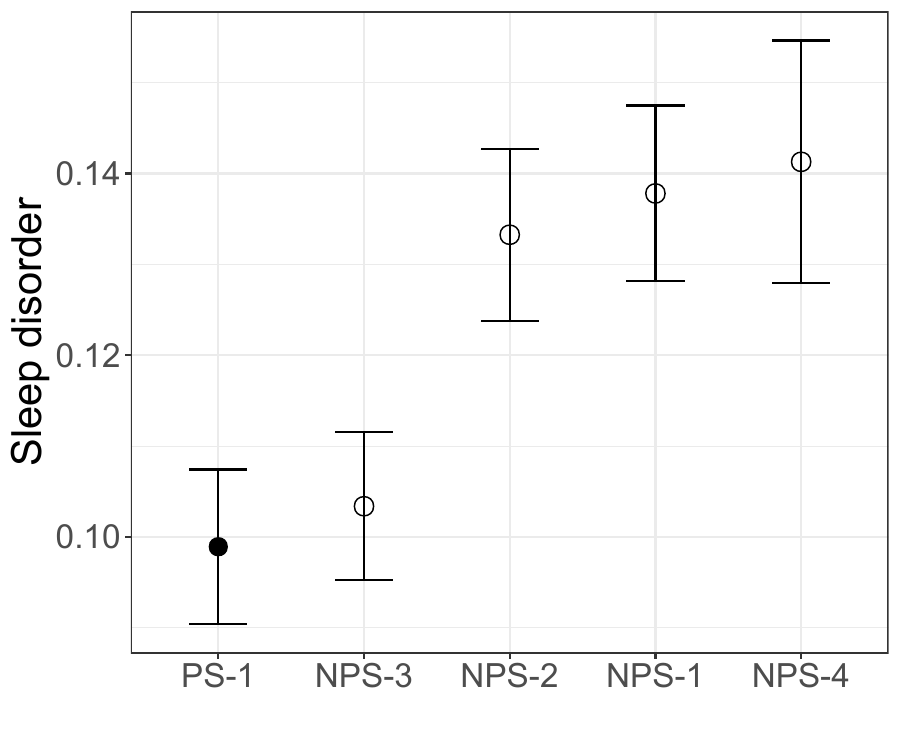}
		\includegraphics[width=0.32\linewidth]{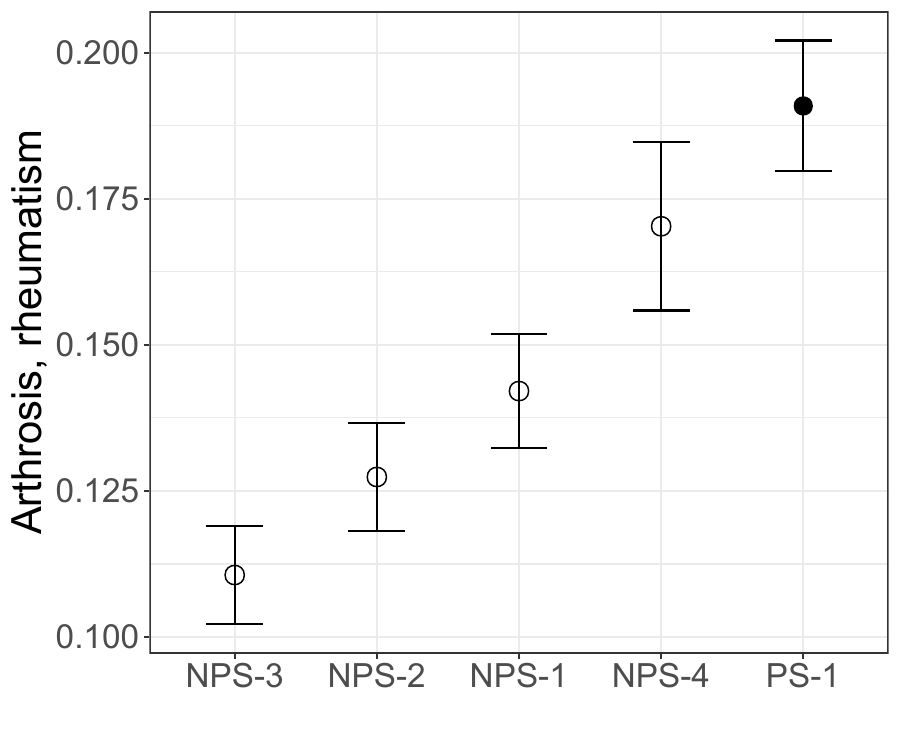}
		\includegraphics[width=0.32\linewidth]{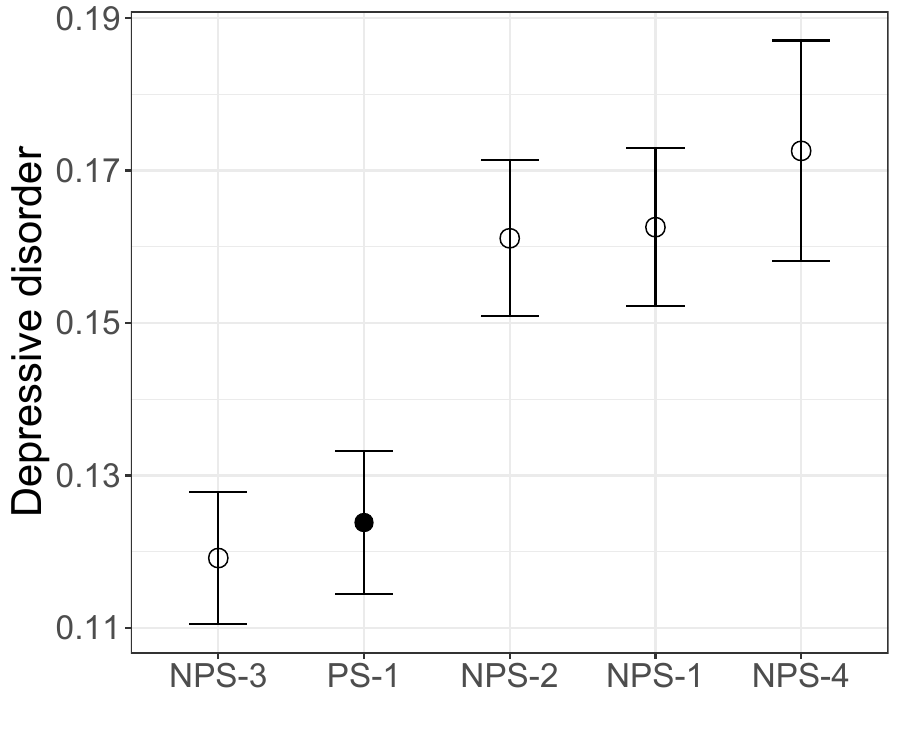}
		\includegraphics[width=0.32\linewidth]{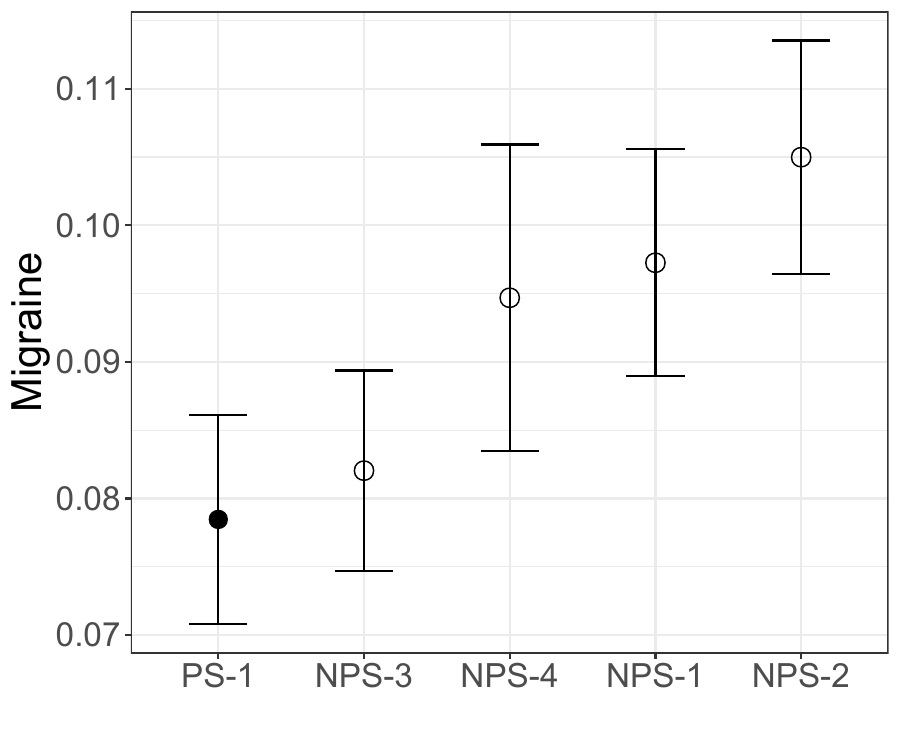}
		\includegraphics[width=0.32\linewidth]{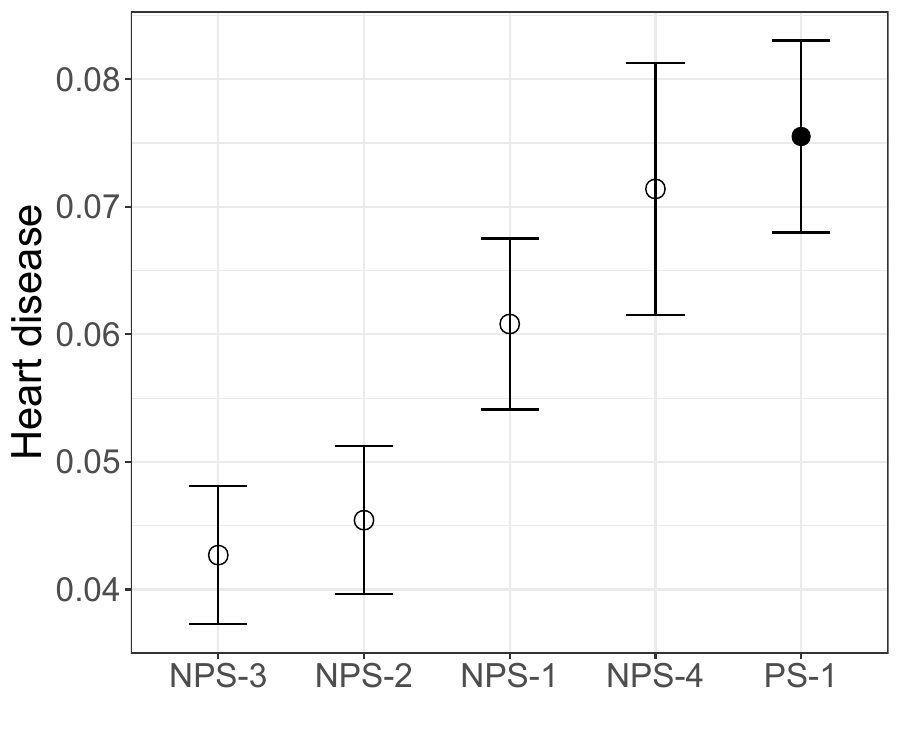}
		\includegraphics[width=0.32\linewidth]{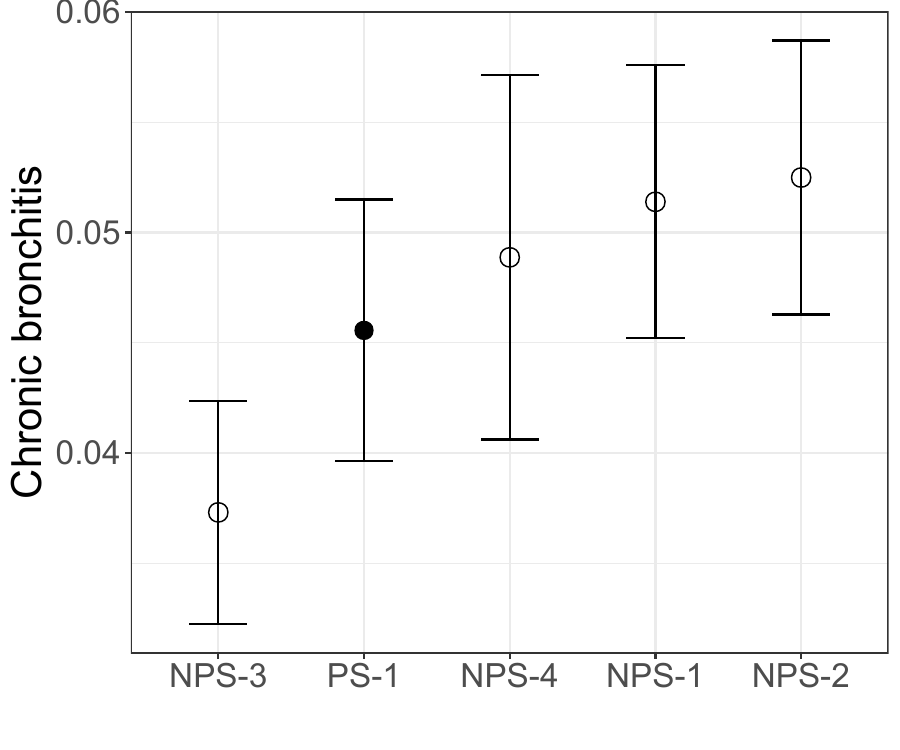}
		\includegraphics[width=0.32\linewidth]{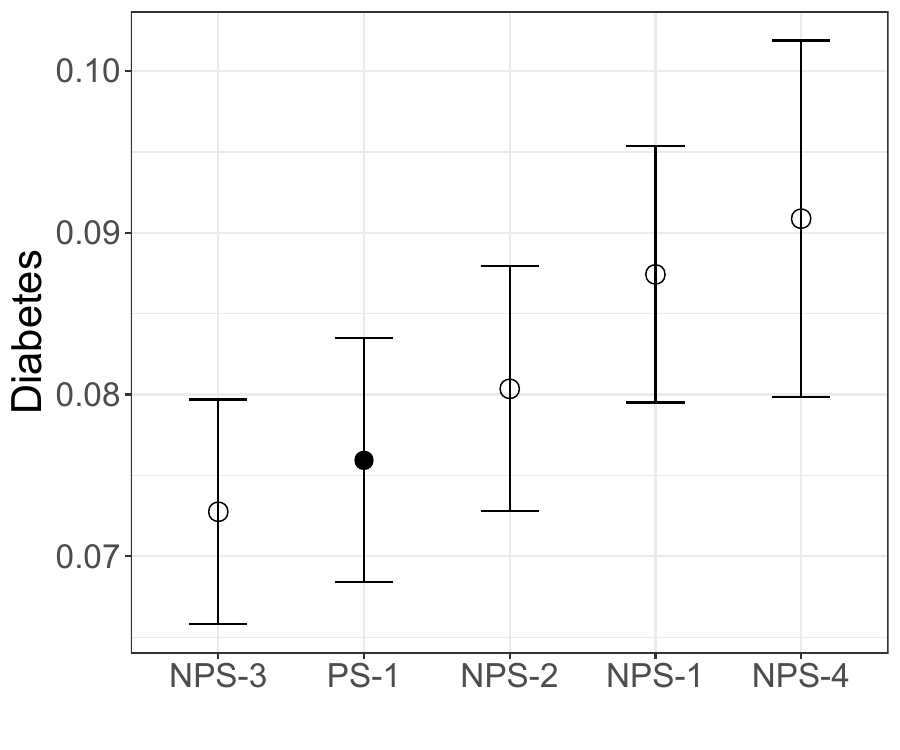}
		\includegraphics[width=0.32\linewidth]{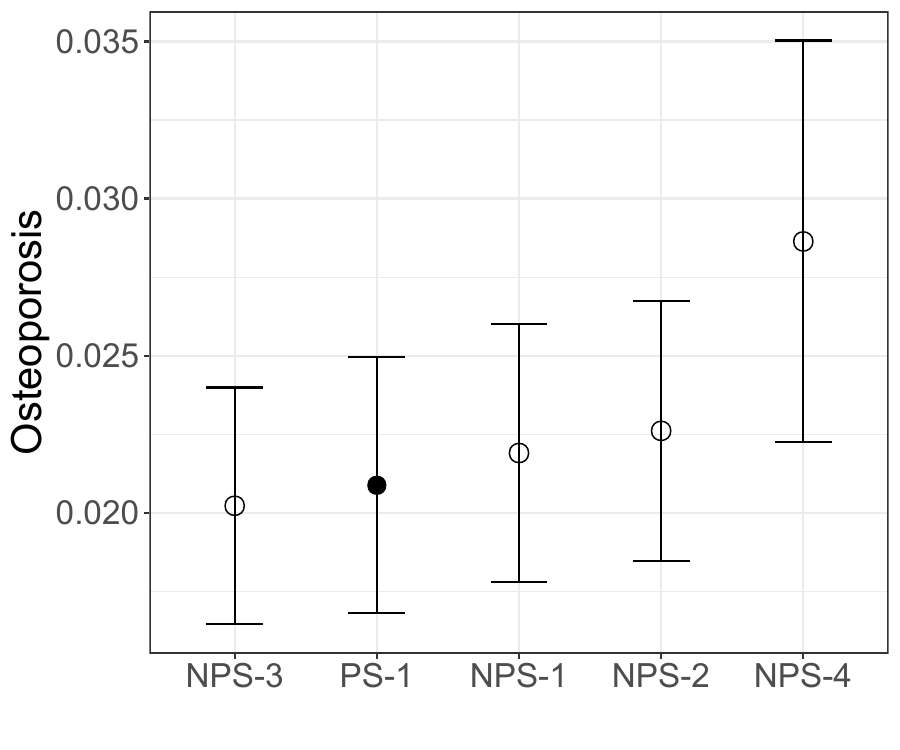}
		\includegraphics[width=0.32\linewidth]{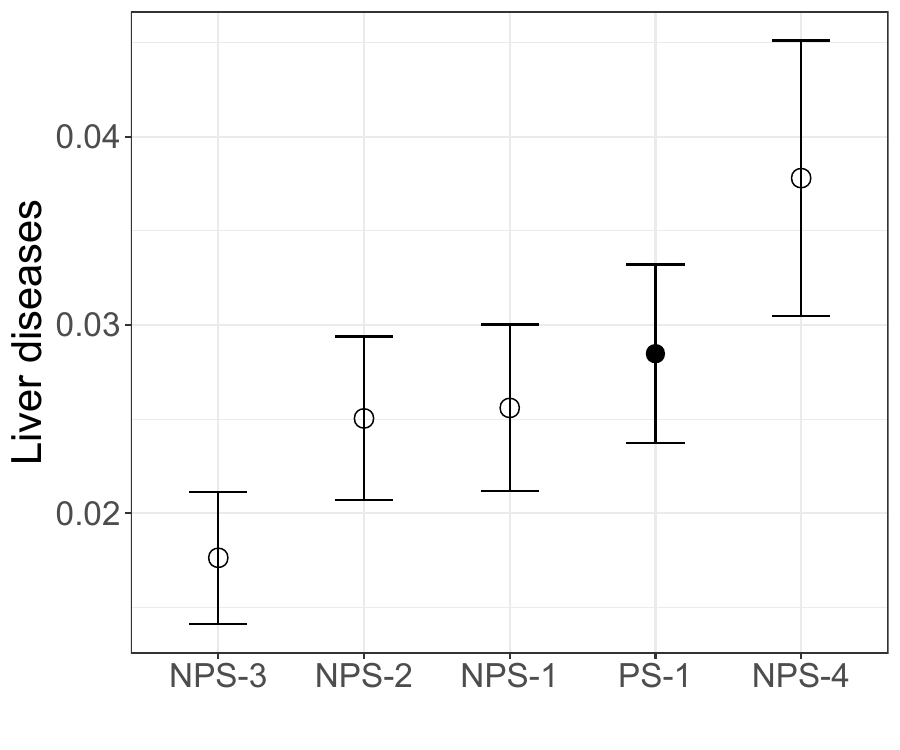}
		\includegraphics[width=0.32\linewidth]{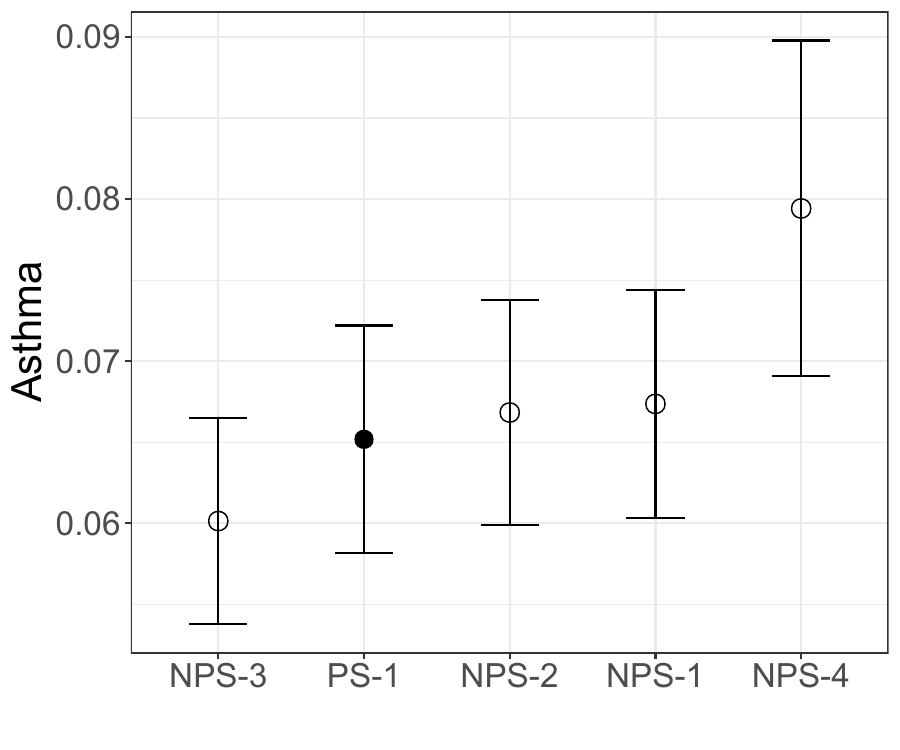}
		\includegraphics[width=0.32\linewidth]{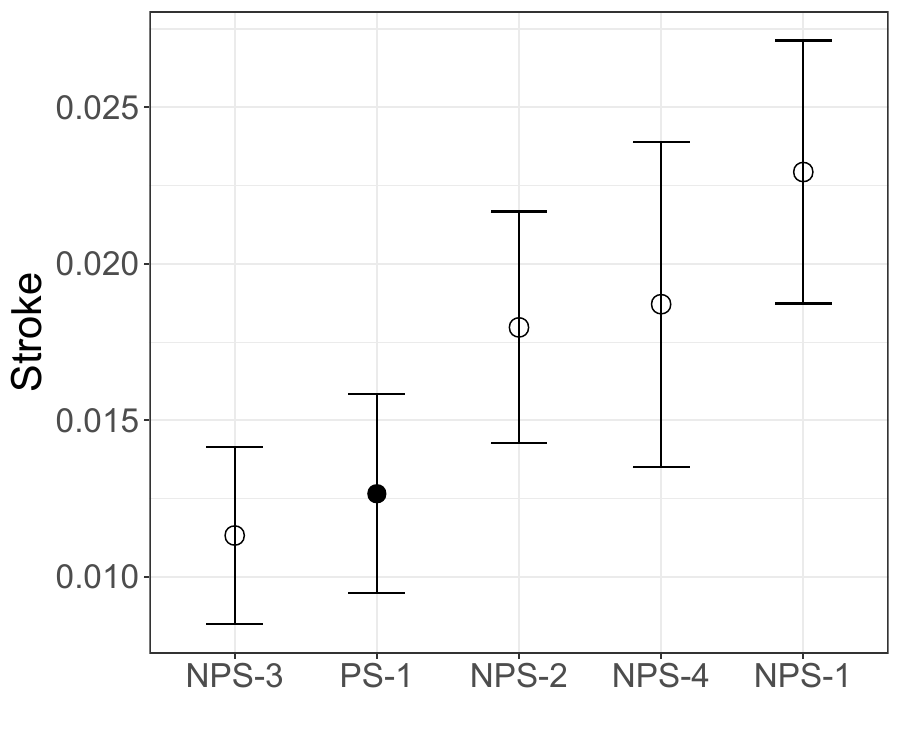}
		\includegraphics[width=0.32\linewidth]{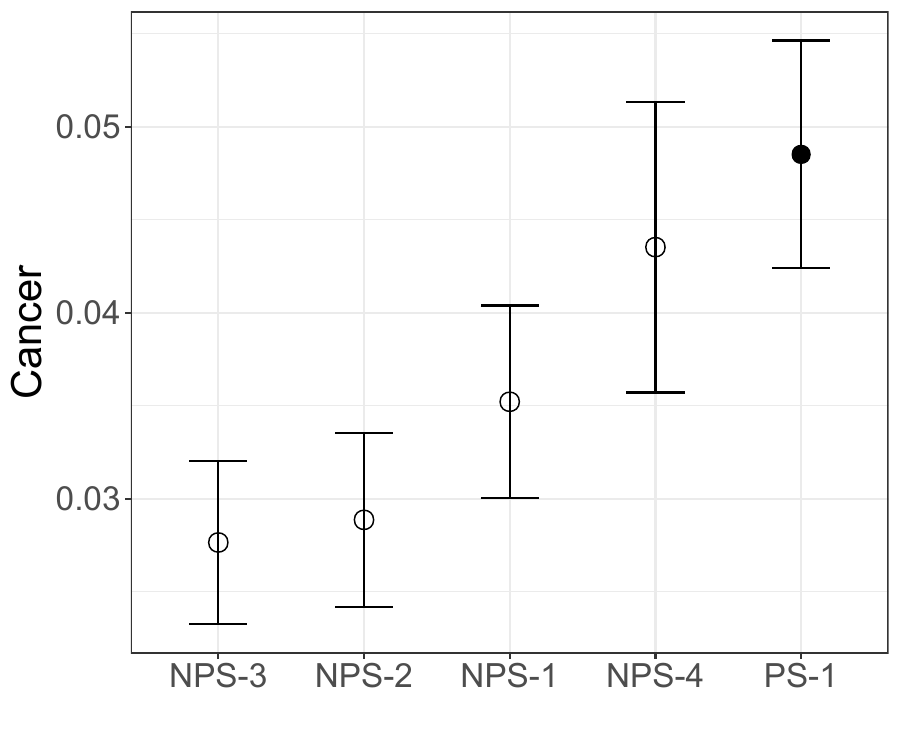}
		\includegraphics[width=0.32\linewidth]{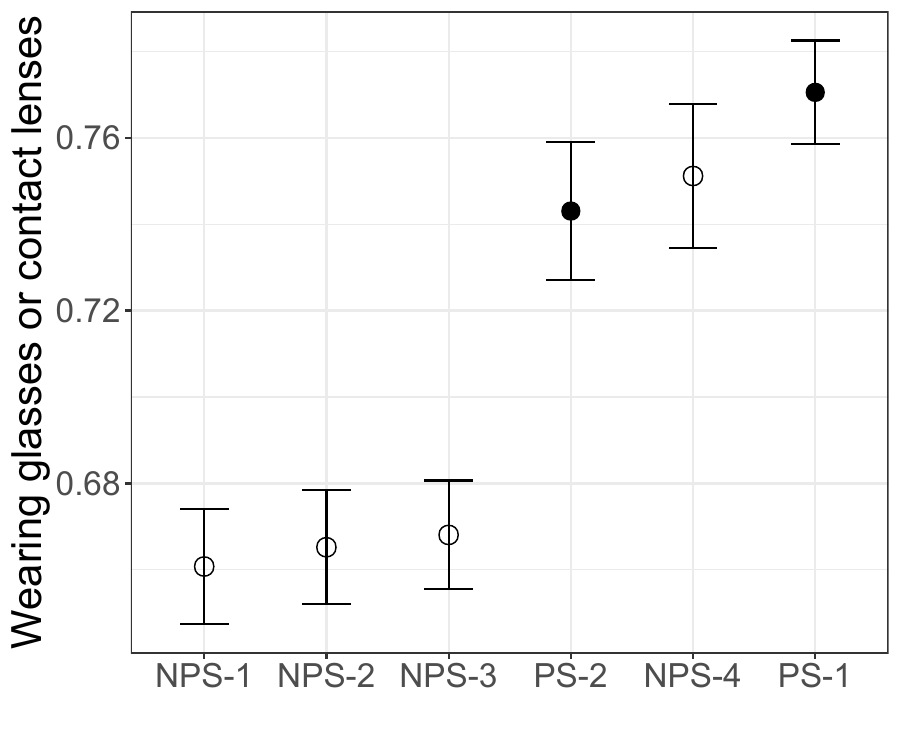}
		\includegraphics[width=0.32\linewidth]{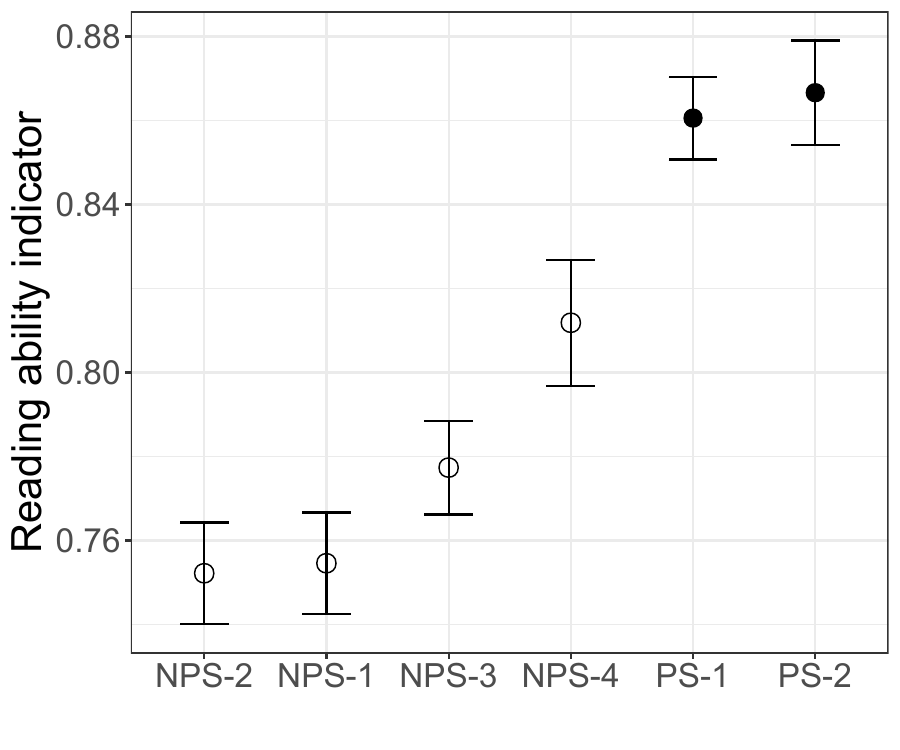}
		\includegraphics[width=0.32\linewidth]{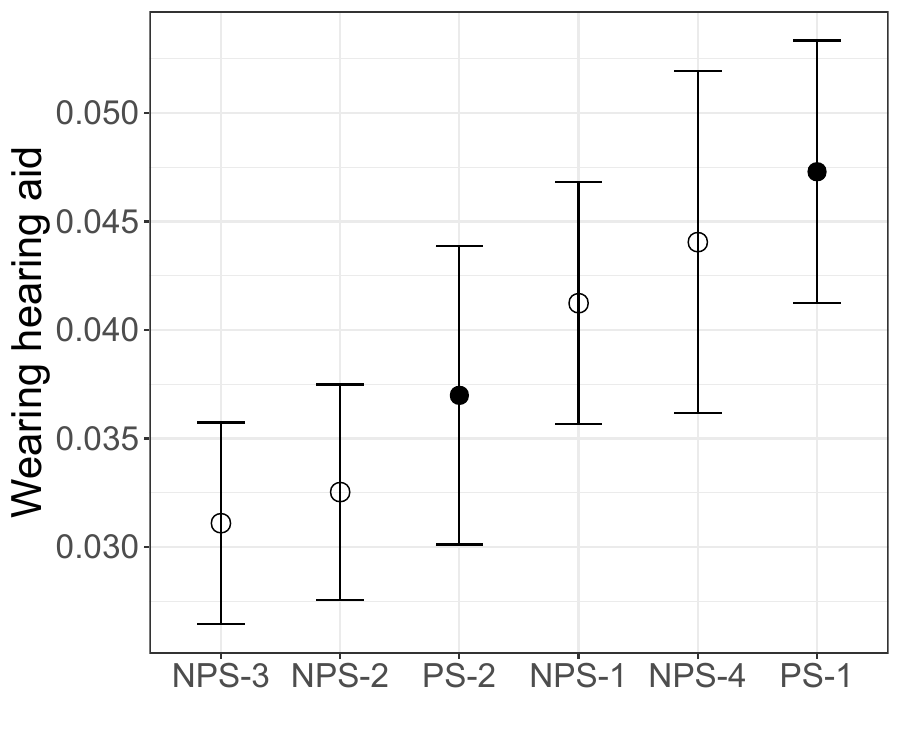}
	\end{subfigure}
\end{figure}
\begin{figure}[htbp]
	\ContinuedFloat
    \centering	
	\begin{subfigure}[htbp]{1\textwidth}
		\caption*{\textit{Figure 2 continued}}
		\includegraphics[width=0.32\linewidth]{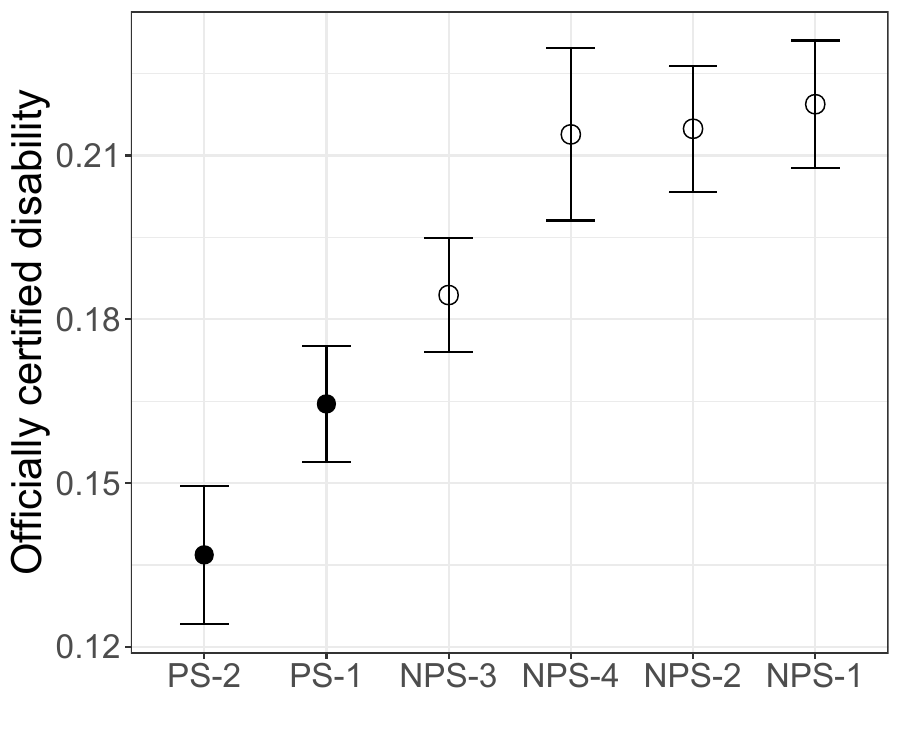}
        \includegraphics[width=0.32\linewidth]{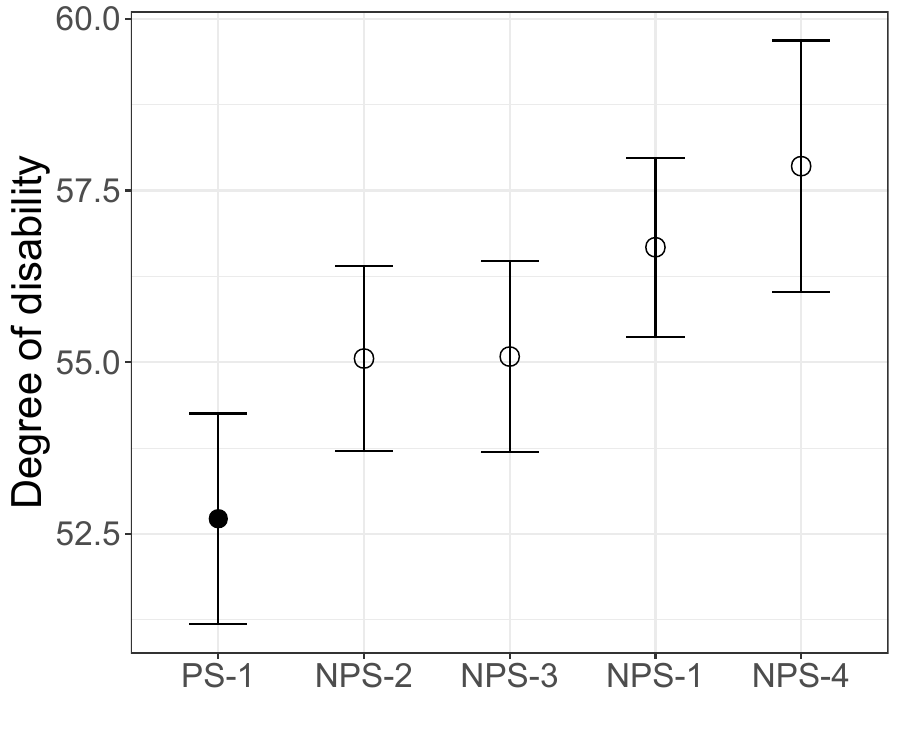}
        \includegraphics[width=0.32\linewidth]{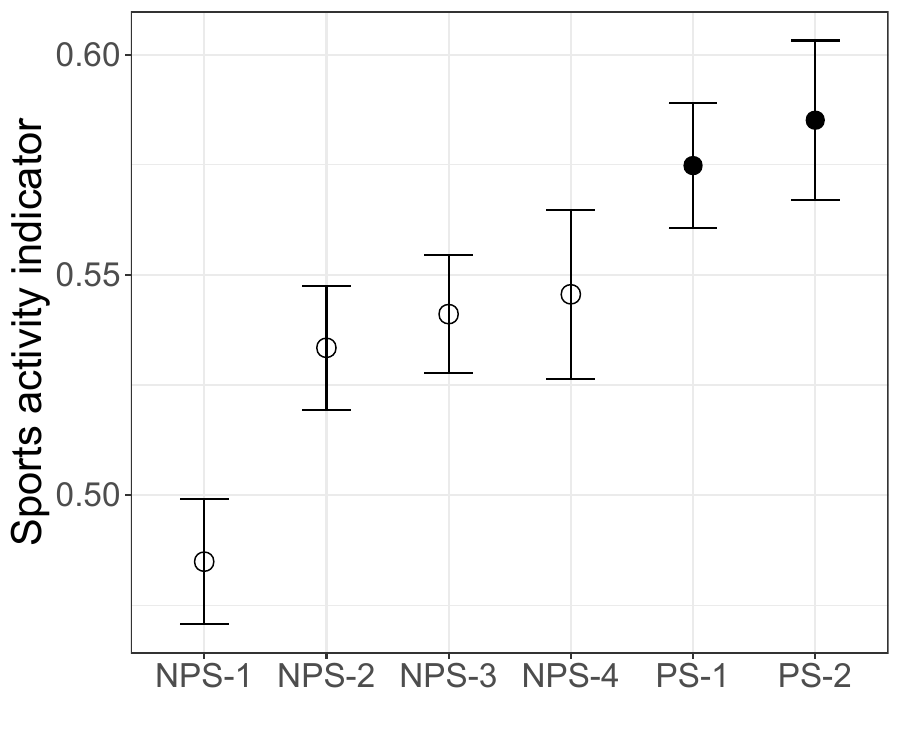}
        \includegraphics[width=0.32\linewidth]{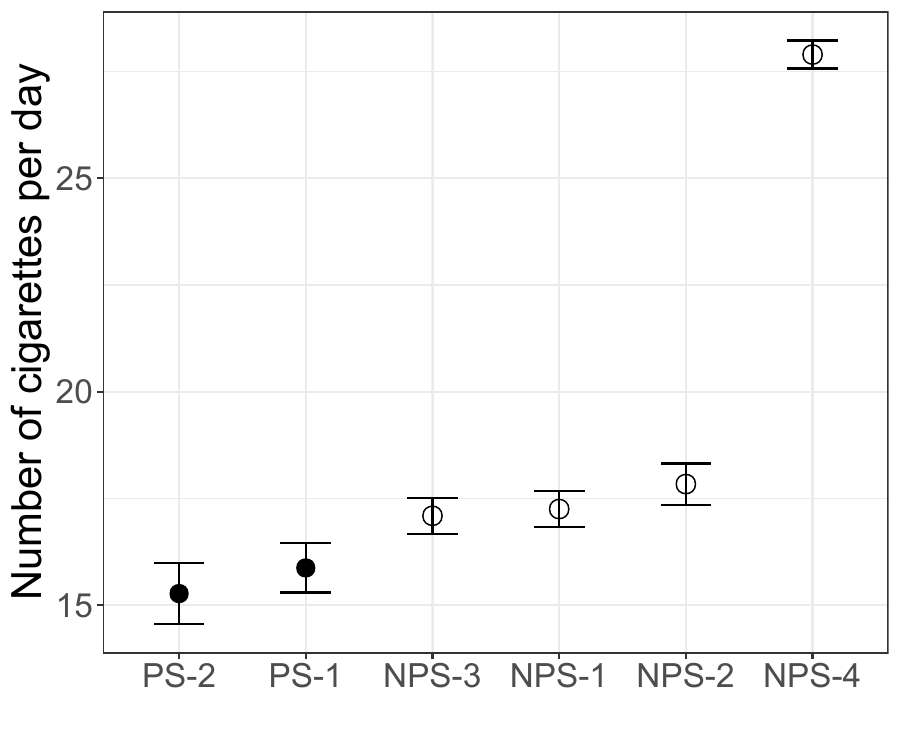}
        \includegraphics[width=0.32\linewidth]{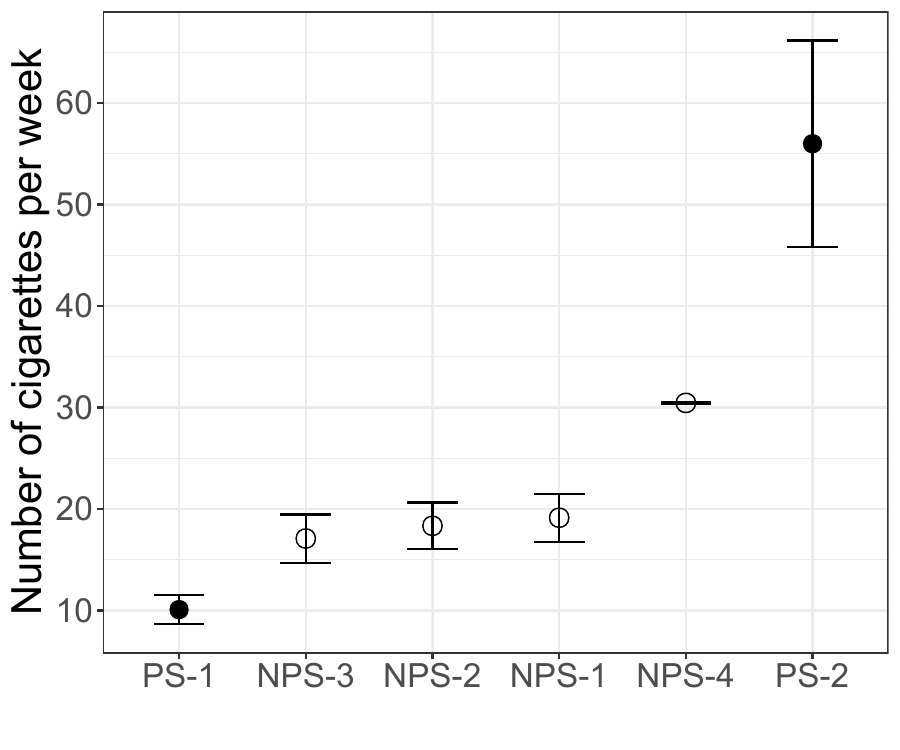}
        \includegraphics[width=0.32\linewidth]{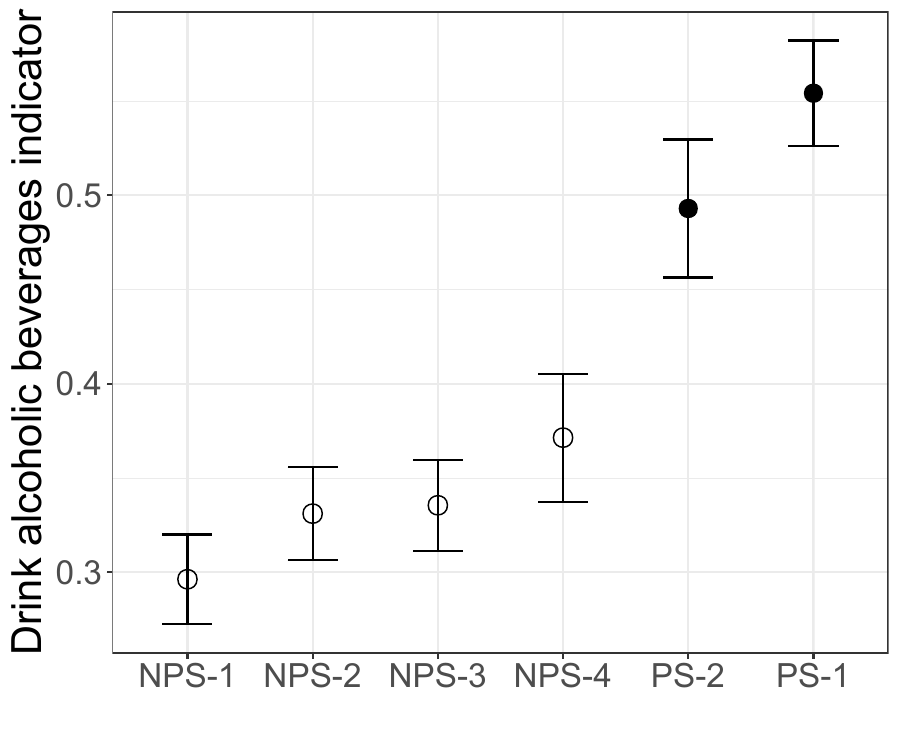}
		\centering
	\end{subfigure}
\caption{Unweighted survey estimates and 95\%-confidence intervals for each survey. The order of surveys on the x-axis of each subgraph corresponds to the size of the estimate. Empty dots indicate non-probability samples; filled dots indicate probability samples.}
	\label{fig:naiveestimatesANOVA}
\end{figure}

\newpage

Separate ANOVAs for each variable showed significant group differences between survey agencies for 34 out of 36 reported variables (94\%). Non-significant ($p > 0.05)$ differences are found for `the number of overnight hospital stays for inpatient treatment within the last 12 months' and `ever diagnosed with osteoporosis'.

Table \ref{tab:pvalues} shows the count of Bonferroni adjusted p-values of t-tests according to conventional significance levels. For the unweighted estimates, 19 variables (56\%) were found to differ significantly between PS-1 and NPS. For PS-2 versus NPS, 8 variables (57\%) showed significant differences.

\begin{table}[htbp]
\centering
\caption{Number of Bonferroni adjusted p-values of t-tests below conventional thresholds. Reading example for the first two lines: (17+2) t-tests corresponds to p-values less or equal 0.05 after correcting for multiple tests. After weighting, (8+4) t-tests still show p-values less or equal 0.05.}
\label{tab:pvalues}
\begin{tabular}{lrrrr}
\toprule
Comparison & $p \leq .001$ & $p \leq .01$ & $p \leq .05$ & $p > .05$\\
\midrule
PS-1 vs NPS (unweighted) & 17 & 0 & 2 & 17\\
PS-1 vs NPS (weighted)   & 8 &  4 & 0 & 24\\
PS-2 vs NPS (unweighted) & 8 &  0 & 0 & 6\\
PS-2 vs NPS (weighted)   & 6 &  1 & 0 & 7\\
\bottomrule
\end{tabular}
\end{table}

The mean effect sizes $d$ \cite{Cohen1988} in the unweighted comparisons are about $d=0.155$ (PS-1 vs NPS) and $d=0.201$ (PS-2 vs NPS). The maxima are $d=0.703$ and $d=0.602$.

To sum up, after accounting for multiple testing, more than half of the variables show significant differences. In addition, the average effect sizes indicate weak effects, but these are larger than, for example, mode effects reported in the literature.\footnote{Effect sizes of mode differences are rarely published in survey methodology. However, \cite{Christmann2009} reports 0.04 as the mean of Cohen's d for 138 items compared between a face-to-face survey and a mixed-mode survey. Compared to these values, the mean effects of NPS vs PS are larger.} However, the strong effect sizes of the maximum values indicate considerable heterogeneity of samples which should reflect the same population.

\subsection{Effects of weighting}

After weighting, Tukey's HSD confirmed remaining significant group differences in 30 out of 36 analyzed health variables (83\%). Due to weighting, significant group differences vanished for four variables (injury at work in the last 12 months, asthma, cancer, and wearing a hearing aid).

Of all 430 pairwise differences, 156 are significant (36\%). Of those, 77 significant group differences were found between the non-probability surveys (49\%), 76 between non-probability and probability surveys (49\%), and three between the two probability surveys (2\%). These three were reading ability, sports activity and the number of smoked cigarettes per week. Denoting the number of differences as $k$, most differences between the non-probability surveys were found between NPS-$4$ and NPS-$2$ ($k=20$), between NPS-$4$ and NPS-$3$ ($k=18$), and least between NPS-$2$ and NPS-$3$ ($k=8$).

In sum, after weighting Tukey’s HSD still showed significant group differences between web surveys in 30 out of 36 analyzed health variables. These remaining differences were not specific to a topic but persisted across question groups (general health status, use of health services, accidents/injuries, disabilities/chronic diseases, and health-related behavior).

As mentioned above in Section \ref{sec:methodscompare}, to reduce the number of comparisons, in addition to the HSD results presented above, we pooled the NPS estimates for the next analysis. Using weighted estimates, comparing PS-1 and NPS yielded 12 variables (36\%) with significant differences. Comparing PS-2 and NPS, 7 variables (50\%) showed significant differences (see Table \ref{tab:pvalues}).  

Moreover, weighting reduced the mean effect sizes to $d=0.063$ and $d=0.127$. The maximum effect sizes after weighting are about $d = 0.255$ (PS-1 vs NPS) and $d = 0.429$ (PS-2 vs NPS), still indicating medium effects, given the traditional classification of effect sizes \cite{Cohen1988}.

To analyze the effect of weighting separately for each item, the relative difference $RD$ between the unweighted $\widehat{Y}_{u}$ and weighted estimates $\widehat{Y}_{w}$ is computed as

\begin{equation}
    RD = \frac{\widehat{Y}_{u} - \widehat{Y}_{w}}{\widehat{Y}_{w}}.
\end{equation}

\begin{figure}[ht]
	\centering
	\includegraphics[width=\textwidth]{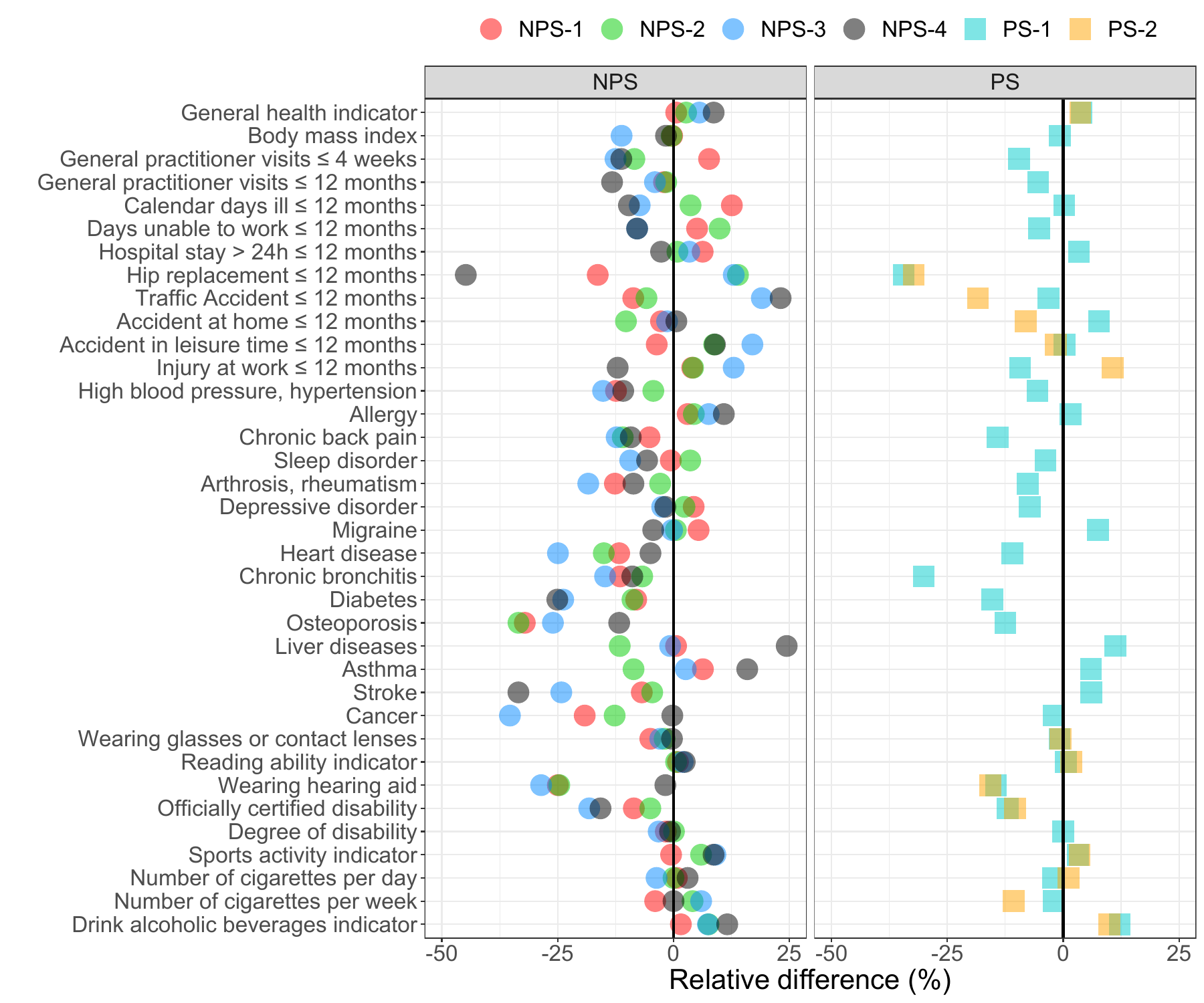}
	\caption{Relative difference between unweighted and weighted survey estimates per survey agency. The black vertical line indicates a zero difference. Y-axis corresponds to the sequence in the questionnaire.}	
    \label{fig:reldiffplot2}
\end{figure}

Hence, a negative $RD$ indicates an underestimation of the unweighted estimate. Figure \ref{fig:reldiffplot2} shows the results. Most comparisons show a negative $RD$ (63\%). Negative differences up to {-45\%} and positive differences up to 25\% are found. In most cases calibration increased the estimates. Hence unweighted estimates would result in underestimations. Therefore, unweighted surveys would overestimate the health of the population. As discussed in section \ref{sec:background}, it is likely that the weighted estimates are still negatively biased.

However, as Figure {\ref{fig:reldiffplot2}} shows, some items such as traffic accidents, liver diseases, or asthma result in large positive differences. Why these specific items produce considerable large overestimations in some NPS and considerable large underestimations in other NPS is unclear. A few outliers are always possible, but the unsystematic pattern, even after calibration, shows that the weighting mechanism does not entirely capture the generating process for the deviations.

\newpage
\section{Summary}

In general, respondents of non-probability samples  were less healthy, used more health services, had more accidents and injuries, and showed more unhealthy behaviour than respondents of the probability samples. Weighting using standard demographics was not capable of removing all significant group differences. About 36\% of the differences between PS-1 and NPS, and 50\% between PS-2 and NPS remained significant after controlling for demographics. Overall, the results show differences between probability and non-probability surveys in health estimates, which were reduced but not eliminated by weighting. Furthermore, the differences between non-probability surveys before and after weighting are larger than expected between random samples from the same population.

\section{Discussion and conclusion}

This study was specifically designed to test for differences between web survey agencies conducting the same study. Weighting reduced some differences in estimates between the surveys, but not all. Therefore, weighting seems insufficient to make estimates of different agencies comparable. Health estimates of non-probability based web surveys rely more on a valid weighting model.

Given that the missing generating mechanism might be MNAR and the still large variation of estimates between different surveys after weighting, collecting health data for population parameter estimates with non-probability based surveys seems not advised.

\section{Limitations}

Although specifically designed, the study presented has certain limitations. Since we wanted to study differences in actual survey practice, the separate effects of non-probability sampling and offline versus online recruitment or different recruitment techniques can not be estimated separately as both factor groups are confounded. However, separating these factors would require the combination of a non-probability sampling with offline recruitment, which we consider unlikely in practice for general populations.

Due to data protection regulations, 22 variables of PS-$2$ could not be used. Therefore, we could not compare all estimates of all surveys. Furthermore, PS-2 was conducted later than the other surveys, which could potentially affect comparisons of this survey with the other surveys. However, neither of these two problems are likely to change the main conclusions. 

Since we used trimmed weights, it could be argued that untrimmed weights might reduce bias. Trimmed weights are used in practice to avoid a large impact of few observations on results, which would increase the sampling variance \cite{HazizaBeaumont2017}. Therefore, trimming weights is common and using a fixed threshold is widespread among commercial survey agencies. 

Since only specific auxiliary variables have been used for calibration, other variables might have reduced bias. However, these auxiliary variables must have been measured, and relevant reference data is needed. For particular topics, such information may be available. For example, \cite{Schonlau2007Dec} used `webographic' variables, and \cite{DiSograCobbChanDennis2011} used early adopter characteristics for weighting. However, since the missing data-generating mechanism for each potential dependent variable could be different, it is unlikely that a universal set of auxiliary variables will be suitable for all variables of interest in a multi-purpose survey. Furthermore, there is no comparative study of other calibration variables than demographics for correcting health bias in web surveys.  

Many different weighting procedures have been suggested in the literature; we considered only the most widely used model in official statistics (calibration). Of course, other models could be applied, for example, multilevel regression with poststratification (MRP) \cite{Gelman1997} or various versions of propensity score adjustment \cite{Rosenbaum1983Apr,Lee2009Feb}. However, these methods will fail if no relevant information on the missing data generating mechanism is contained in the auxiliary information used in the model \cite{Bruch2022, Copas2020Dec, Si2020O, Hanretty2019}. Testing this proposition will be the topic of a follow-up paper.

The selection steps for a web-only health survey were described in section \ref{subsec:internethealth}. It is unclear if the effect of health-related issues on  
participation in health surveys can be explained by  
ICD codes alone. If symptoms and their severeness and not diagnosis are relevant for  survey participation, much more detailed questions than usual within general population surveys are required. Therefore, detailed studies of specific diagnoses and symptoms as factors in answering web surveys seem to be advised.

Only survey estimates were compared, and no external data was used for validation in the study reported here. The details of a validation study comparing weighted and unweighted estimates of the six surveys with external data are subject of ongoing work of the authors. 

\section*{Acknowledgements}

The views expressed in this paper are those of the authors and do not necessarily reflect the policies of their affiliations.

\section*{Funding}
This research was funded by the research grant 286253962 of the German Research Foundation (DFG), granted to the first author.

\section*{Availability of data and materials}

The programs and survey datasets used in the current study are available from the corresponding author upon reasonable request. The administrative datasets supporting this study's findings are  available from German Official Statistics under access restrictions, requiring an individual licence, which was obtained for this study. 

\section*{Ethics approval and consent to participate}
Data was collected online after information on data processing was provided. Informed consent was inferred by the completion of the web survey. There were no experiments on humans, and no human tissue samples were used in this study. The Faculty of Social Sciences Ethics Committee (University of Duisburg-Essen, Decision DFG-286253962) approved the study and all procedures used.

\section*{Competing interests}
The authors declare that they have no competing interests.

\section*{Consent for publication}
Not applicable.

\section*{Authors' contributions}
RS designed the study, wrote the grant proposal, commissioned and supervised the data collection and revised the first manuscript draft written by JK. Data analysis was conceptualized and computed by RS and JK. Furthermore, JK managed the commissioning of the data collection process. 

\bibliographystyle{abbrv}
\bibliography{References}

\end{document}